\documentclass[11pt,a4paper]{article}

\usepackage[margin=23mm]{geometry}
\usepackage[T1]{fontenc}
\usepackage[utf8]{inputenc}
\usepackage{lmodern}
\usepackage{microtype}
\usepackage[authoryear,round]{natbib}
\usepackage{amssymb}
\usepackage{amsmath}
\usepackage[colorlinks=true,allcolors=blue]{hyperref}
\usepackage[nameinlink,noabbrev]{cleveref}
\usepackage{enumitem}
\usepackage{booktabs}
\usepackage{subcaption}
\usepackage{adjustbox}
\usepackage{tocloft}
\usepackage{makecell}
\usepackage{float}
\usepackage{xcolor} 
\usepackage{multirow}
\usepackage{url}
\usepackage{placeins}
\usepackage{caption}

\begin{document}

\title{Improvements to the post-processing of weather forecasts using machine learning and feature selection}
\author{%
Kazuma Iwase$^{1}$ and Tomoyuki Takenawa$^{1}$\thanks{Corresponding author. Email: \href{mailto:takenawa@kaiyodai.ac.jp}{takenawa@kaiyodai.ac.jp}}\\[0.4em]
\small $^{1}$Graduate School of Marine Science and Technology, Tokyo University of Marine Science and Technology,\\
\small 2-1-6 Etchujima, Koto-ku, Tokyo 135-8533, Japan}
\date{}

\maketitle

\begin{abstract}
This study aims to develop and improve machine learning-based post-processing models for precipitation, temperature, and wind speed predictions using the Mesoscale Model (MSM) dataset provided by the Japan Meteorological Agency (JMA) for 18 locations across Japan, including plains, mountainous regions, and islands. 
By incorporating meteorological variables from grid points surrounding the target locations as input features and applying feature selection based on correlation analysis, we found that, in our experimental setting, the LightGBM-based models achieved lower RMSE than the specific neural-network baselines tested in this study, including a reproduced CNN baseline, and also generally achieved lower RMSE than both the raw MSM forecasts and the JMA post-processing product, MSM Guidance (MSMG), across many locations and forecast lead times. 
Because precipitation has a highly skewed distribution with many zero cases, we additionally examined Tweedie-based loss functions and event-weighted training strategies for precipitation forecasting. These improved event-oriented performance relative to the original LightGBM model, especially at higher rainfall thresholds, although the gains were site dependent and overall performance remained slightly below MSMG.
\end{abstract}

\noindent\textbf{Keywords:} Weather prediction; post-processing; machine learning; feature selection

\section{Introduction}
\label{sec:introduction}

\subsection{Research Background and Objectives}
\label{subsec:research_background_and_objectives}

Numerical weather prediction (NWP) is a primary method of weather forecasting that represents future atmospheric states using mathematical models based on observation data. Over the years, it has evolved through advancements in the representation of physical processes, model initialization, including assimilation, and the introduction of ensemble modeling \citep{bauer2015quiet}. However, NWP is not without its limitations, such as forecast errors and biases. While forecast variance might be unavoidable, several post-processing techniques using machine learning models have been developed to mitigate these biases
\citep{liu2023deep,rojas2023postprocessing,yoshikane2022bias,zhang2021machine,kudo2022statistical,peng2020prediction,tang2021numerical,salazar2022multivariable,xu2020wind}.

The Japan Meteorological Agency (JMA) has also introduced a post-processing method called the MSM Guidance (MSMG) to correct the errors of the Mesoscale Model (MSM), which has a finer grid spacing (5 km) and targets areas around Japan compared to global models \citep{JMA2024}. In MSMG, JMA primarily employs statistical methods, such as the Kalman filter, frequency bias correction, and neural networks, to reduce the systematic biases in MSM.

This study aims to develop machine learning-based post-processing models for precipitation, temperature, and wind speed predictions—representative meteorological variables—using MSM data for 18 locations across Japan, including plains, mountainous regions, and islands. Meteorological variables from grid points surrounding the prediction locations were used as input features, and feature selection based on correlation analysis was applied. 
In our experimental setting, the LightGBM-based models achieved lower RMSE than the tested neural-network baselines, including the reproduced CNN baseline, and also generally achieved lower RMSE than both the raw MSM forecasts and the JMA post-processing product, MSM Guidance (MSMG). Among the LightGBM-based models, those using surrounding-grid information with correlation-based feature selection showed the lowest RMSE across many locations and forecast lead times.

For precipitation, we further examined Tweedie-based loss functions and event-weighted training strategies, which improved event-oriented metrics for some sites and rainfall thresholds even when the gains in RMSE were limited.

\subsection{Related Work}
\label{subsec:related_work}

Several studies have developed post-processing models for precipitation using methods such as convolutional neural networks (CNNs), neural networks, and Support Vector Machines \citep{zhang2021machine,liu2023deep,rojas2023postprocessing,yoshikane2022bias}. Among them, \cite{zhang2021machine} conducted comparative experiments on machine learning models, input parameters, and training data periods for precipitation prediction, and concluded that LightGBM provides the most balanced performance among the tested models.

For temperature post-processing models, CNNs, neural networks, and LightGBM have been employed \citep{kudo2022statistical,peng2020prediction,tang2021numerical}. \cite{kudo2022statistical} developed a CNN-based temperature post-processing model targeting the Kanto region of Japan and demonstrated its superior performance over the JMA's MSMG.

Wind speed post-processing models have also been explored using neural networks and LightGBM \citep{salazar2022multivariable,tang2021numerical,xu2020wind}. \cite{xu2020wind} utilized LightGBM for wind speed prediction, analyzing feature importance derived from the model. Their study revealed that using all features, including various weather elements, as input outperformed models where features were limited primarily to wind speed.

In addition to using NWP outputs, some studies have leveraged meteorological forecast data from existing weather forecasting services \citep{iwase2024interpolation,tsipis2023improving}. \cite{iwase2024interpolation} employed machine learning models such as LightGBM, XGBoost, and neural networks to predict temperature and precipitation in mountainous regions. By incorporating surrounding weather forecast data as input variables, the results were superior to existing weather forecast services.

In recent machine learning research, deep learning methods have been developed significantly in fields such as natural language processing and image recognition. In contrast, for tabular data, tree-based models currently outperform deep learning models in terms of accuracy and require less hyperparameter tuning costs \citep{shwartz2022tabular,grinsztajn2022tree}. The utility of tree-based models has also been demonstrated in post-processing NWP outputs, which often involve tabular data \citep{zhang2021machine,xu2020wind}. Recently, \citet{hieta2025operational} reported that a tree-based gradient-boosting post-processing scheme (XGBoost) reduces RMSE for short-range near-surface forecasts.

\vskip\baselineskip
The remainder of this paper is organized as follows. \Cref{sec:methods} defines the methods and data used in this study. \Cref{sec:results_and_discussion} presents the results and discussion. Finally, \Cref{sec:conclusion} concludes the study.

\section{Methods}
\label{sec:methods}

\subsection{Data}
\label{subsec:data}
\begin{figure}[!htbp]
\centering
\includegraphics[width=0.82\textwidth]{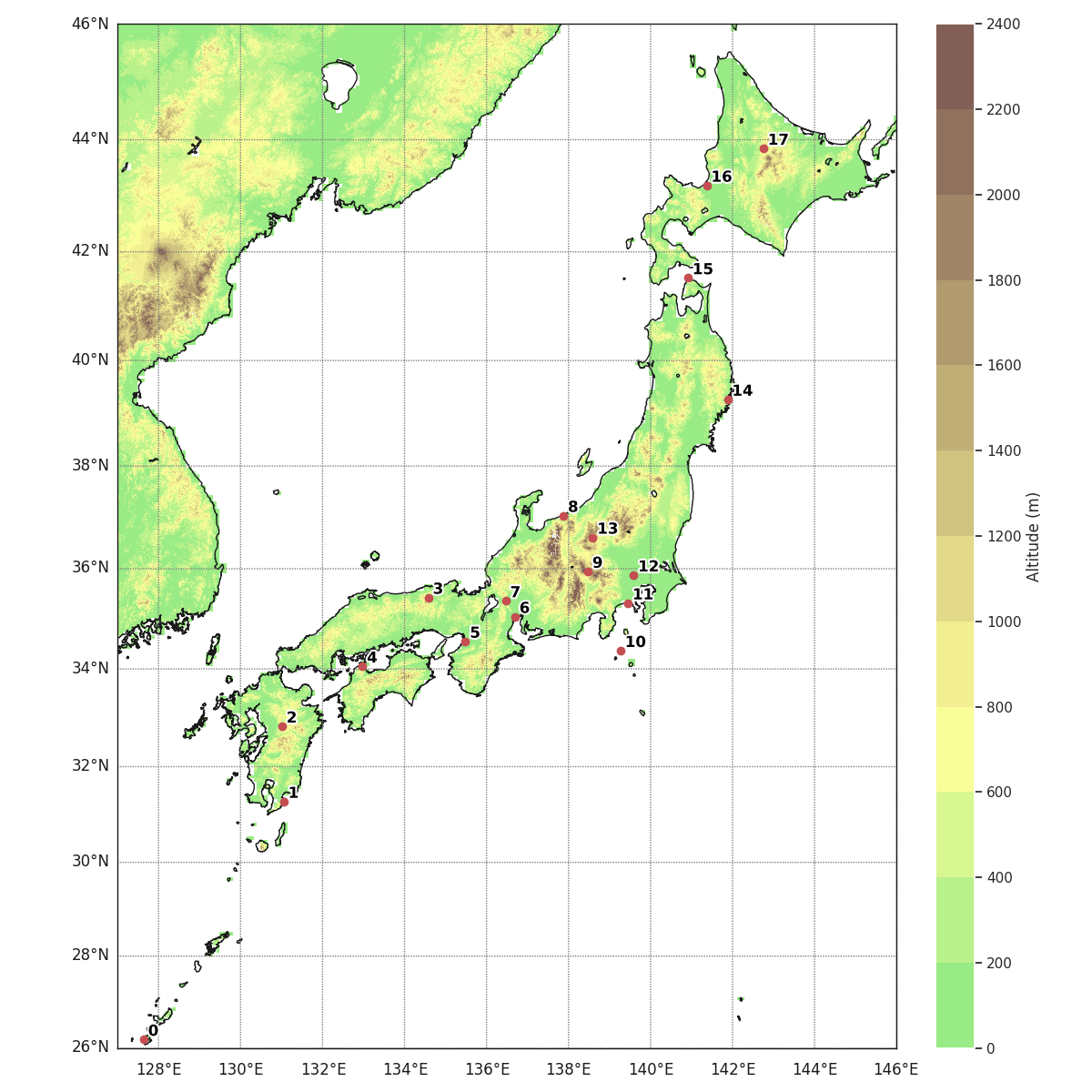}
\caption{Map showing the 18 selected locations.}
\label{fig:map}
\end{figure}

In this study, we use MSM data from the JMA as the input for the model and observation data as the target data for training. The MSM data were collected and distributed by Research Institute for Sustainable Humanosphere, Kyoto University \citep{RISH2024}. The observation data from JMA include cumulative precipitation over the past three hours, hourly temperature, and wind speed. We selected the 18 observation sites from all over Japan (\Cref{fig:map}, \Cref{tab:pos_info}). Additionally, in order to compare the forecast results, we collected MSMG data, a post-processed forecast provided by JMA and distributed by the Japan Meteorological Business Support Center (JMBSC).

\begin{table}[!htbp]
\caption{Latitude, longitude, altitude, and station numbers of the 18 sites shown in \Cref{fig:map}.}
\label{tab:pos_info}
\begin{adjustbox}{center}
\begin{tabular}{l|rrrr}
\toprule
 & Latitude & Longitude & Altitude & Station Number \\
\midrule
Ashimine & 26.193& 127.638& 3 & 0 \\
Uchinoura & 31.277& 131.055& 8 & 1 \\
Minamiaso & 32.832& 131.013& 394 & 2 \\
Uwanokogen & 35.432& 134.583& 540 & 3 \\
Imabari & 34.053& 132.975& 27 & 4 \\
Sakai & 34.555& 135.485& 20 & 5 \\
Kuwana & 35.050& 136.693& 3 & 6 \\
Sekigahara & 35.363& 136.467& 130 & 7 \\
Itoigawa & 37.043& 137.875& 8 & 8 \\
Nobeyama & 35.948& 138.472& 1350 & 9 \\
Niijima & 34.368& 139.268& 29 & 10 \\
Tsujido & 35.320& 139.450& 5 & 11 \\
Saitama & 35.875& 139.587& 8 & 12 \\
Kusatsu & 36.617& 138.592& 1223 & 13 \\
Kamaishi & 39.270& 141.878& 5 & 14 \\
Oma & 41.527& 140.912& 14 & 15 \\
Ishikari & 43.193& 141.370& 5 & 16 \\
Kamikawa & 43.847& 142.753& 324 & 17 \\
\bottomrule
\end{tabular}
\end{adjustbox}
\end{table}

MSM consists of three-dimensional gridded predictions of future atmospheric conditions, including temperature, wind, moisture, and solar radiation, with fine grid spacing of approximately 5 km. The forecasts cover Japan and surrounding sea areas \citep{JMA2024}. Surface-level data are available at hourly intervals, while pressure-level data are available at three-hour intervals. The forecasts are issued every three hours (00, 03, 06, 09, 12, 15, 18, 21 UTC) and extend up to 39 hours ahead. The surface grid comprises 505 × 481 points (grid spacing: 0.05° latitude × 0.0625° longitude), while the pressure-level grid consists of 253 × 241 points (grid spacing: 0.1° latitude × 0.125° longitude), covering the area from the northwest corner (47.6°N, 120°E) to the southeast corner (22.4°N, 150°E).

MSM outputs a variety of meteorological elements on the surface and on multiple standard pressure levels.
In this study, we use the elements listed in \Cref{tab:meteo_element} as predictors.
For pressure-level variables, although MSM provides outputs at more pressure levels, we use only three representative levels (850, 500, and 200~hPa), following prior work \citep{liu2023deep}, to capture information from the lower, middle, and upper atmosphere while reducing dimensionality.
The CNN baseline uses a different input configuration following prior work (including additional near-surface pressure levels); see Section~\ref{subsec:Neural_Network_and_CNN}.

\begin{table}[!htbp]
\centering
\caption{Meteorological elements used as predictors, including names and units.
A $\checkmark$ indicates that the element is used at the surface and/or at the representative pressure levels (850, 500, and 200~hPa).
(Relative humidity is used at 850/500~hPa only.)}
\label{tab:meteo_element}
\begin{tabular}{llll}
\toprule
\textbf{Element Name}    & \textbf{Surface} & \textbf{Pressure Level} & \textbf{Unit} \\
 								& (hourly) & (every 3 hours) &\\ 
\midrule
Temperature             & \checkmark       & \checkmark             & $^\circ$C            \\
Relative humidity       & \checkmark       &850/500 hPa only     & \%            \\
Hourly precipitation    & \checkmark       &                        & mm            \\
Wind speed              & \checkmark       & \checkmark             & m s$^{-1}$           \\
U wind component        & \checkmark       & \checkmark             & m s$^{-1}$           \\
V wind component        & \checkmark       & \checkmark             & m s$^{-1}$           \\
Vertical velocity       &                  & \checkmark             & hPa s$^{-1}$       \\
Surface pressure        & \checkmark       &                        & hPa           \\
Sea level pressure      & \checkmark       &                        & hPa           \\
Geopotential height     &                  & \checkmark             & m              \\
Solar radiation         & \checkmark       &                        & W m$^{-2}$     \\
Total cloud cover       & \checkmark       &                        & \%            \\
Low cloud cover         & \checkmark       &                        & \%            \\
Mid-level cloud cover   & \checkmark       &                        & \%            \\
High cloud cover        & \checkmark       &                        & \%            \\
\bottomrule
\end{tabular}
\end{table}



For each station, we train an independent regression model.
Except for the CNN-based model, the inputs to LightGBM and the neural network
are one-dimensional feature vectors constructed by concatenating
(i) surface variables within a temporal window $(t, t-1, t-2)$,
(ii) pressure-level variables at the target time $t$ only (850, 500, and 200~hPa),
and (iii) auxiliary scalar features (lead time, month, and hour of day).
All gridded predictors are flattened before concatenation.
The temporal window length was chosen to match the 3-hourly forecast cycle of MSM/MSMG,
so that a consistent set of predictors is available across lead times and forecast cycles.

The input variables consist of 13 surface variables and 7 pressure-level variables
as summarized in \Cref{tab:meteo_element}.
Among the pressure-level variables, relative humidity is available only at two levels
(850 and 500~hPa), whereas the other six variables are available at all three levels.

We prepared two input configurations.
In the first, only the grid cell closest to the prediction point is used.
In this configuration, the total number of input features is 62, computed as
\[
  13 \times 3 + (6 \times 3 + 2) + 3 = 62,
\]
where $13 \times 3$ accounts for the 13 surface variables at three time steps,
$(6 \times 3 + 2)$ counts pressure-level combinations
(six variables at three levels plus relative humidity at two levels) at time $t$,
and $+3$ accounts for lead time, month, and hour of day.

In the second, surrounding grid cells are included to assess the benefit of spatial context.
Because the pressure-level fields have twice the horizontal grid spacing of the surface fields,
we use an $11 \times 11$ grid for surface variables and a $7 \times 7$ grid for pressure-level variables
so that the covered areas are comparable.
Under this surrounding-grid configuration, the total number of input features is 5,702, computed as
\[
  13 \times 3 \times 11 \times 11 + (6 \times 3 + 2) \times 7 \times 7 + 3 = 5702,
\]
where $13$ is the number of surface variables, $3$ is the number of time steps for surface variables,
$11 \times 11$ is the surface grid, $(6 \times 3 + 2)$ counts pressure-level
combinations at time $t$, $7 \times 7$ is the pressure-level grid, and $+3$ accounts for lead time, month, and hour of day.

\begin{figure}[!htbp]
    \centering
    \begin{subfigure}{\textwidth}
        \centering
        \includegraphics[width=0.9\linewidth]{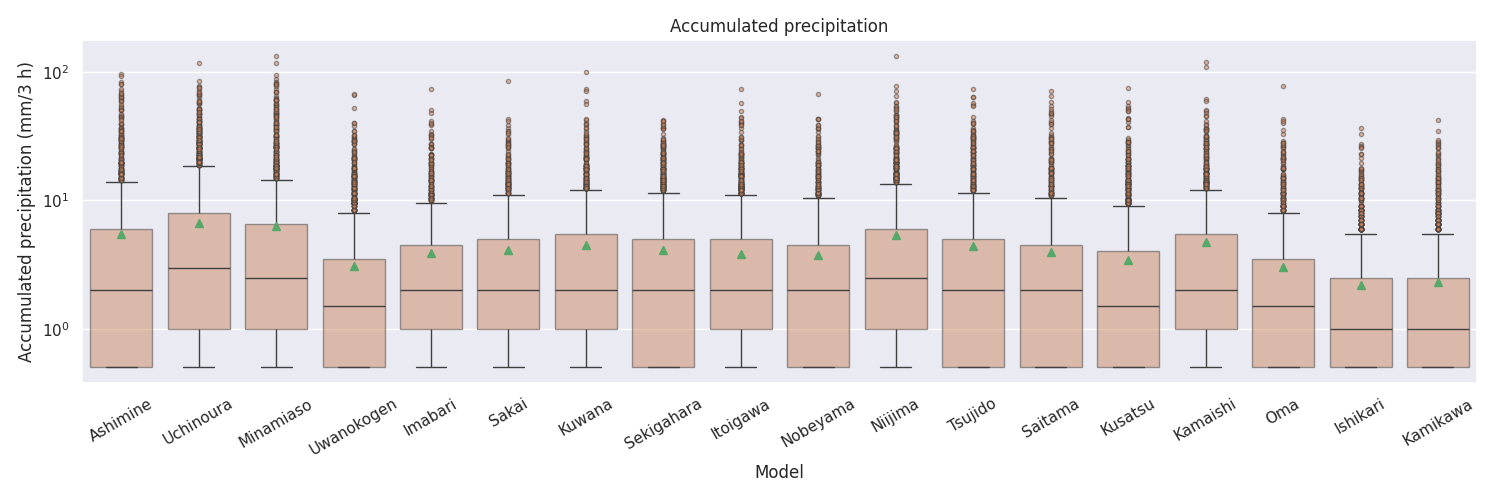}
        \caption{Precipitation}
    \end{subfigure}
    \vspace{0.4em}

    \begin{subfigure}{\textwidth}
        \centering
        \includegraphics[width=0.9\linewidth]{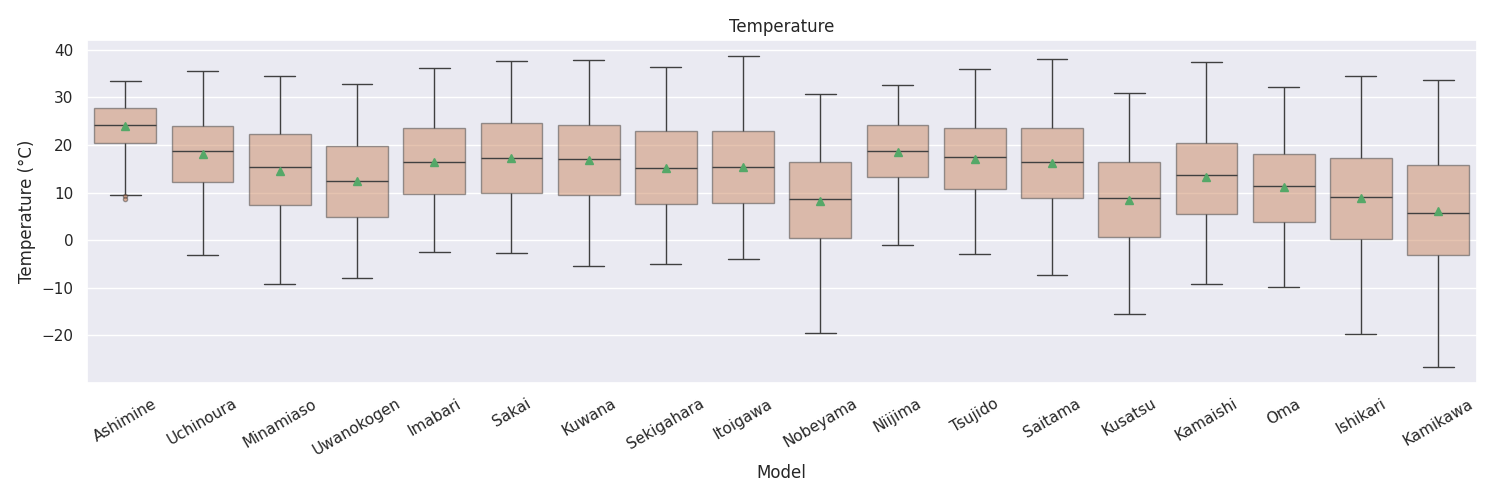}
        \caption{Temperature}
    \end{subfigure}
    \vspace{0.4em}

    \begin{subfigure}{\textwidth}
        \centering
        \includegraphics[width=0.9\linewidth]{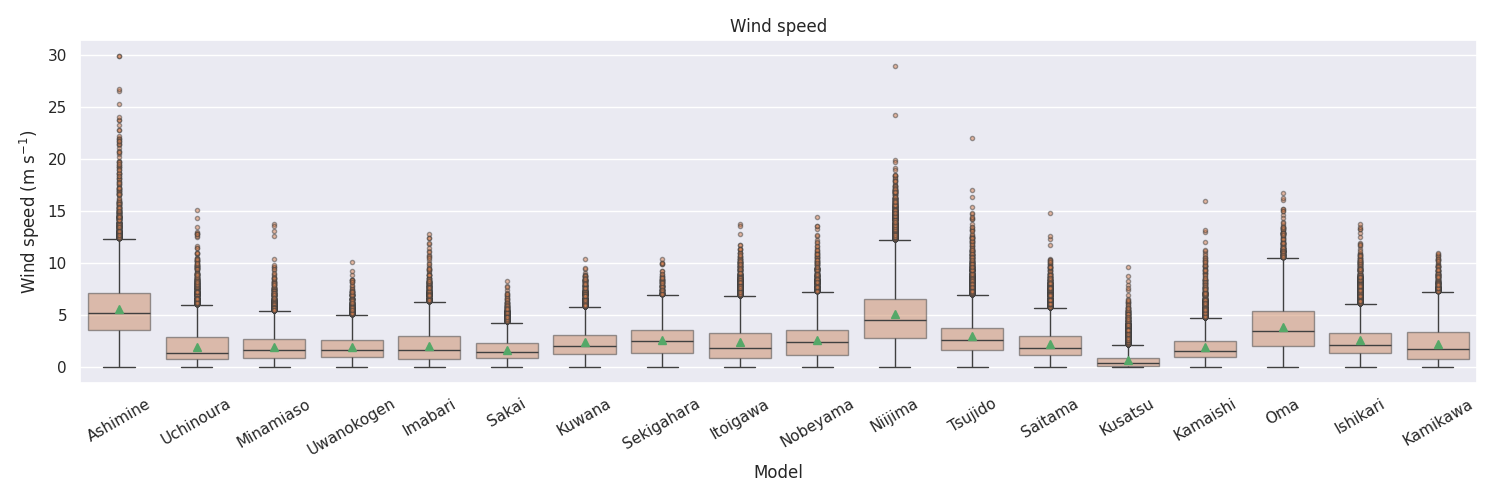}
        \caption{Wind Speed}
    \end{subfigure}
    
\caption{Box plots of the distributions of each meteorological variable at each location. For precipitation, only nonzero values are shown (values $> 0$~mm) to avoid the dominance of zero precipitation and to provide an informative visualization; the precipitation axis is displayed on a logarithmic scale. The green triangle markers indicate the mean values.}
\label{fig:box_plot}
\end{figure}

The output variables are the cumulative precipitation over the past three hours, as well as hourly temperature and wind speed. Observation data from 2019 to 2021 were used for training, 2022 for validation, and 2023 for testing, as shown in \Cref{tab:data_size}. Since the MSM provides complete fields without missing values, only observation samples containing missing values were excluded from the dataset.
The validation data were used solely for model selection and 
hyperparameter tuning, and were not used to retrain the model prior to testing.  \Cref{fig:box_plot} presents box plots for each location and meteorological variable.

\begin{table}[!htbp]
\caption{The number of samples in the training, validation, and test datasets, and the total number of features when using the surrounding grids of the prediction point, as well as the number of features after feature selection based on correlation analysis.}
\label{tab:data_size}
\resizebox{\textwidth}{!}{
\begin{tabular}{l|rrrrr}
\toprule
 & Train samples & Validation samples & Test samples & Original Features & Dropped features \\
\midrule
Ashimine & 113854 & 37882 & 37934 & 5702 & 239 \\
Uchinoura & 113854 & 37843 & 37947 & 5702 & 296 \\
Minamiaso & 113828 & 37830 & 37960 & 5702 & 502 \\
Uwanokogen & 108642 & 37570 & 37947 & 5702 & 385 \\
Imabari & 113802 & 37830 & 37934 & 5702 & 408 \\
Sakai & 113776 & 37830 & 37947 & 5702 & 455 \\
Kuwana & 111891 & 37817 & 37960 & 5702 & 354 \\
Sekigahara & 113568 & 37830 & 37934 & 5702 & 437 \\
Itoigawa & 113609 & 37843 & 37817 & 5702 & 516 \\
Nobeyama & 113022 & 37544 & 37791 & 5702 & 869 \\
Niijima & 113477 & 37778 & 37921 & 5702 & 236 \\
Tsujido & 113763 & 37791 & 37960 & 5702 & 416 \\
Saitama & 113789 & 37817 & 37960 & 5702 & 248 \\
Kusatsu & 112359 & 37271 & 37752 & 5702 & 733 \\
Kamaishi & 113464 & 37687 & 37921 & 5702 & 295 \\
Oma & 113815 & 37804 & 37895 & 5702 & 256 \\
Ishikari & 113243 & 37791 & 37934 & 5702 & 271 \\
Kamikawa & 113607 & 36907 & 37791 & 5702 & 413 \\
\bottomrule
\end{tabular}
}
\end{table}

For MSMG, the accumulated surface precipitation over the past three hours and hourly surface temperature and wind speed are predicted up to 39 hours ahead, with forecast outputs available every three hours (00, 03, 06, 09, 12, 15, 18, and 21 UTC). Precipitation forecasts are provided on a grid with dimensions of 560 (longitude) × 480 (latitude), corresponding to a grid spacing of 0.05$^\circ$ in latitude and 0.0625$^\circ$ in longitude, covering the area from (47.975$^\circ$N, 120.03125$^\circ$E) in the northwest to (20.025$^\circ$N, 149.96875$^\circ$E) in the southeast. In contrast, forecasts for temperature and wind speed are provided at the locations of JMA observation stations.

The dataset is summarized as follows. The outputs consist of three variables
(precipitation, temperature, and wind speed), while the inputs are identical
across them. Missing values occur only in the observation data. 
The period covers the entire span of the training (2019--2021), validation (2022), and test
(2023) sets. The total number of rows is determined by the combination of period,
forecast cycles, lead times, and locations, while the number of columns differs
according to the input configuration.

\begin{description}
  \item[Period:] 2019--2023 (1,826 days)
  \item[Forecast cycles per day:] 8 (3-hourly)
  \item[Lead times per cycle:] 13 (3--39 h)
  \item[Locations:] 18
  \item[Total rows:] 3,418,272 (before removing missing observations)
\end{description}

\subsection{LightGBM}
\label{subsec:LightGBM}
Several previous studies on tabular data and post-processing model development have reported that decision tree-based models outperform other machine learning models in terms of training cost and performance scores. Among decision tree-based models, LightGBM \citep{ke2017lightgbm} is particularly fast and high-performing \citep{iwase2024interpolation}. Therefore, this study adopts LightGBM for the development of the post-processing model. LightGBM is an ensemble model based on decision trees, employing a boosting algorithm that sequentially adds decision trees and uses a quadratic optimization method. 

Since LightGBM cannot handle multiple output variables simultaneously, a separate model should be created for each output variable. During training, Early Stopping was applied using validation data with a patience of 10 epochs, and the loss function was set to Mean Squared Error (MSE).
\begin{equation}
\text{MSE} = \frac{1}{N} \sum_{i=1}^{N} \left( \hat{y}_i - y_i \right)^2
\end{equation}
where $N$ is the number of data points, $\hat{y}_i$ is the predicted value, and $y_i$ is the observed value.

\subsection{Feature Selection}
\label{subsec:feature_selection}
When surrounding grids are included, the number of features reached 5,702, which is excessive for LightGBM. Feature selection is necessary to reduce training time and prevent overfitting. To achieve this, the correlation analysis method proposed in \citep{bedi2018attribute}, denoted as FS0 hereafter, was applied. 
FS0 computes pairwise absolute Pearson correlations among candidate features and removes redundancies.

To avoid removing critical input information, we exclude features at the grid closest to the prediction target and calendar variables from the analysis by dropping columns. Thus, the correlation analysis is performed solely among features from surrounding grids.

Let $\tau$ denote the correlation threshold. FS0 builds the absolute correlation matrix $[|r(x_i,x_j)|]_{i,j}$ and scans features in their column order. For each feature $x_j$, if there exists any earlier feature $x_i$ $(i<j)$ with $|r(x_i,x_j)|>\tau$, then $x_j$ is dropped (left-to-right rule). Based on validation data, the threshold was set to $\tau=0.9$. The number of features after correlation analysis for each location is summarized in \Cref{tab:data_size}.

After applying the aforementioned feature selection process or in the case where only single-grid data were used, the following three further feature selection methods were applied to determine the optimal number of features:
\begin{enumerate}[label=FS\arabic*.]
    \item A method based on the correlation coefficient between each feature and the output variable.
    \item A method based on the mutual information between each feature and the output variable.
    \item A method based on feature importance calculated by LightGBM.
\end{enumerate}

While many feature selection methods exist, this study adopted representative approaches \citep{chandrashekar2014survey,ross2014mutual}. The correlation coefficient-based method calculates the correlation coefficient between each feature and the output variable, and selects features in descending order of the correlation coefficient. The mutual information-based method uses mutual information between each feature and the output variable, where mutual information measures the degree of dependence between two variables \citep{ross2014mutual}. The LightGBM feature importance-based method calculates the contribution of each feature to reducing the loss function during tree splitting in LightGBM. Additionally, an analysis of the number of selected features was conducted. Variables such as prediction time, month, and hour were included in the final selected features regardless of the feature selection results.

\subsection{Parameter Tuning}
\label{subsec:Parameter_tuning}
Parameter tuning was conducted using Optuna \citep{akiba2019optuna}, a parameter optimization library based on Bayesian optimization algorithms. Specifically, this study employed LightGBMTuner \citep{LightGBMTuner} to compute optimal values for each parameter iteratively. The tuning process was performed using a single representative location, and the tuned parameters were applied to other locations. The chosen representative location was Sekigahara, situated near the center of all target locations. Since LightGBM cannot handle multiple output variables, tuning was conducted separately for each output variable.

\subsection{Neural Network and CNN}
\label{subsec:Neural_Network_and_CNN} 
To provide comparisons with neural-network-based approaches used in previous studies, we also implemented a feedforward neural network and a CNN model. The structure of the neural network, implemented using TensorFlow \citep{TensorFlow} with keras.layers, is as follows:
\begin{itemize}
    \item Dense(1000, activation='relu')
    \item Dense(500, activation='relu')
    \item Dense(100, activation='relu')
    \item Dense(3)
\end{itemize}
The input dimension corresponds to the number of features after FS0 selection, 
and the output dimension (3) corresponds to precipitation, temperature, and wind speed.

The model was trained with the Adam optimizer (learning rate 0.001) \citep{kingma2014adam}, 
MSE as the loss, and Root Mean Squared Error (RMSE) :
\begin{equation}
\text{RMSE} = \sqrt{\frac{1}{N} \sum_{i=1}^{N} \left( \hat{y}_i - y_i \right)^2} \label{eq:rmse}
\end{equation}
as a metric.
Training used early stopping with a patience of 10 epochs, based on validation RMSE.
We set a maximum of 1000 epochs with a batch size of 64; 
in practice, training typically converged within about 25–50 epochs due to early stopping.

CNN models have many possible variations, including the number of layers, kernel sizes, numbers of channels, the use of skip connections, and the choice of optimization and regularization methods.
A comprehensive comparison of these architectures is beyond the scope of the present paper.
Accordingly, we limited the comparison to a CNN model based on prior work that used CNNs for post-processing MSM data.
For the CNN baseline, we followed the experimental setting of \citet{kudo2022statistical} to reproduce a representative prior CNN-based approach under comparable data conditions.
Accordingly, the CNN was applied only to temperature prediction, as in the original study.

\begin{figure}[!htbp]
\centering
\includegraphics[width=0.78\textwidth]{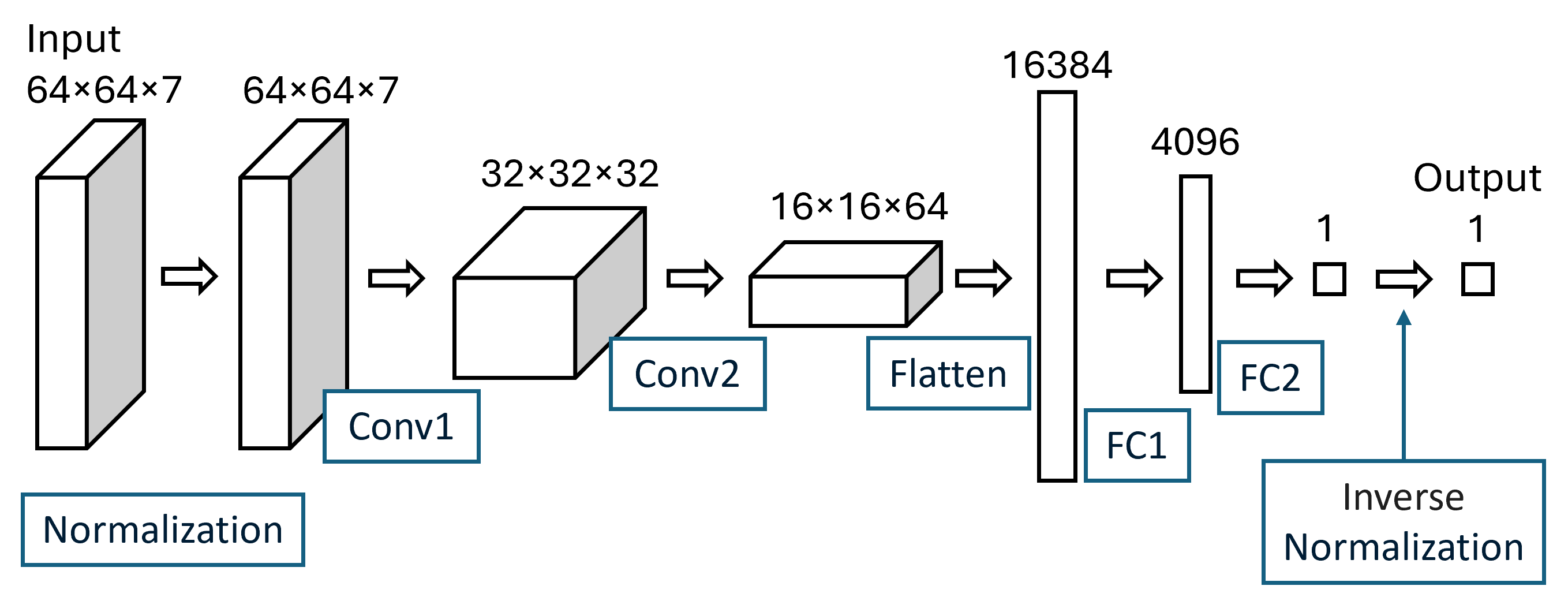}
\caption{Architecture of the CNN.}
\label{fig:cnn}
\end{figure}

Our CNN uses a single forecast-time spatial field without a lookback window.
The input is a $64\times 64\times 7$ tensor centered at each of the 18 locations.
The seven channels consist of four surface-level fields (surface temperature, mean sea-level pressure, and surface wind components $U$ and $V$) and temperature at three pressure levels (975, 925, and 850~hPa).
Because the pressure-level grid is coarser than the surface grid, the pressure-level temperature fields are first cropped as $32\times 32$ patches on the pressure-level grid and then upsampled to $64\times 64$ to match the surface fields.
The architecture of the CNN is shown in \Cref{fig:cnn} and summarized in \Cref{tab:cnn_structure}.

Normalization for CNN was performed as follows:
\begin{equation}
x' = \frac{x - x_{\text{min}}}{x_{\text{max}} - x_{\text{min}}}
\end{equation}
where $x'$ represents the normalized value, $x$ is the original value, and $x_{\text{min}}$ and $x_{\text{max}}$ are the minimum and maximum values of the MSM input data. These minimum and maximum values were calculated for each prediction time (data sample) rather than across the entire training dataset. For temperature, $x_{\text{min}}$ was adjusted to $x_{\text{min}} - 3$ and $x_{\text{max}}$ to $x_{\text{max}} + 3$. An inverse normalization step was applied after the output layer, defined as:
\begin{equation}
y' = y \cdot (x_{\text{max}} - x_{\text{min}}) + x_{\text{min}}
\end{equation}
where $y$ is the output from the output layer, and $y'$ is the value after inverse normalization.
These normalization and inverse-normalization procedures were introduced to reproduce the prior study as closely as possible, and should thus be understood as part of the baseline implementation rather than as independently optimized design choices in this study.

The learning settings for CNN, including the optimizer, loss function, and batch size, were the same as for NN. 
We set a maximum of 1000 epochs; in practice, training typically converged within about 15–20 epochs due to early stopping.

\begin{table}[!htbp]
\centering
\caption{Summary of processing at each CNN layer. Each layer's implementation is described using TensorFlow \citep{TensorFlow} keras.layers.}
\label{tab:cnn_structure}
\begin{tabular}{ll}
\toprule
\textbf{Layer} & \makecell{\textbf{Function}} \\
\midrule
               & ZeroPadding2D(padding=(2, 2)) \\
               & Conv2D(32, kernel\_size=5, strides=1, padding='valid') \\
\textbf{Conv1} & MaxPooling2D(pool\_size=2, strides=2) \\
               & BatchNormalization() \\
               & Activation('relu') \\
\midrule
               & ZeroPadding2D(padding=(2, 2)) \\
               & Conv2D(64, kernel\_size=5, strides=1, padding='valid') \\
\textbf{Conv2} & MaxPooling2D(pool\_size=2, strides=2) \\
               & BatchNormalization() \\
               & Activation('relu') \\
\midrule
               & Dense(4096) \\
\textbf{FC1}   & BatchNormalization() \\
               & Activation('relu') \\
\midrule
\textbf{FC2}   & Dense(1) \\
\bottomrule
\end{tabular}
\end{table}

\subsection{Evaluation Metrics}
\label{subsec:evaluation_metrics}

In addition to RMSE, we used Mean Error (ME),
\[
ME = \frac{1}{N}\sum_{i=1}^{N}(\hat{y}_i-y_i),
\]
to assess model bias, where \(N\) is the number of data points, \(\hat{y}_i\) is the predicted value, and \(y_i\) is the observed value.

In addition, the following metrics were used to further assess model performance for specific variables.

\subsubsection{Statistical Tests for Temperature and Wind Speed}
\label{subsubsec:statistical_tests}

Statistical comparisons were conducted using paired absolute errors. Let
\[
e_i^{\mathrm{main}} = \left|\hat{y}_i^{\mathrm{main}} - y_i\right|,
\qquad
e_i^{\mathrm{base}} = \left|\hat{y}_i^{\mathrm{base}} - y_i\right|
\]
denote the absolute errors of the main and baseline models for sample \(i\).
Then the mean absolute errors (MAEs) are defined as
\[
\mathrm{MAE}_{\mathrm{main}} =
\frac{1}{N}\sum_{i=1}^{N} e_i^{\mathrm{main}},
\qquad
\mathrm{MAE}_{\mathrm{base}} =
\frac{1}{N}\sum_{i=1}^{N} e_i^{\mathrm{base}},
\]
and the paired differences are defined by
\[
d_i = e_i^{\mathrm{main}} - e_i^{\mathrm{base}}.
\]
Based on these paired differences, we conducted a paired \(t\)-test and a
Wilcoxon signed-rank test for temperature and wind speed. Because of the
very large sample size, these tests are interpreted only as supplementary
information, and practical significance is assessed mainly from the error
metrics themselves.

\subsubsection{Event-Based Metrics for Precipitation}
\label{subsubsec:event_metrics_precip}

For precipitation, forecast skill was additionally evaluated using event-based metrics.
For each precipitation threshold \(\tau\), forecasts and observations were converted into binary events.
Let \(H(\tau)\), \(M(\tau)\), and \(F(\tau)\) denote the numbers of hits, misses, and false alarms, respectively, at threshold \(\tau\).
Then, the event-based metrics were defined as
\[
\begin{aligned}
\mathrm{TS}(\tau)   &= \frac{H(\tau)}{H(\tau)+M(\tau)+F(\tau)}, \\
\mathrm{POD}(\tau)  &= \frac{H(\tau)}{H(\tau)+M(\tau)}, \\
\mathrm{FAR}(\tau)  &= \frac{F(\tau)}{H(\tau)+F(\tau)}, \\
\mathrm{Bias}(\tau) &= \frac{H(\tau)+F(\tau)}{H(\tau)+M(\tau)}.
\end{aligned}
\]

\subsection{Weighted Tweedie Model for Precipitation}
\label{subsec:weighted_tweedie}

Precipitation is a nonnegative variable with a highly skewed distribution and many zero or near-zero values.
Under such conditions, standard regression training based on mean squared error tends to be dominated by the large number of weak-rainfall or no-rainfall samples, and may therefore lead to predictions biased toward intermediate values, without sufficiently improving forecast accuracy for moderate to heavy precipitation events.
To address these characteristics, in this study we examined a model that uses a Tweedie loss function with sample weight coefficients in LightGBM training for precipitation prediction (hereafter referred to as the weighted Tweedie model) \citep{jorgensen1997,ke2017lightgbm,dunn2005}.
In preliminary experiments for a few sites, the weighted Tweedie model showed more improvements than the variants using only weighting or only the Tweedie loss, so we focus on this variant here.

Let $y_i \ge 0$ denote the observed precipitation for sample $i$, and let $\mu_i > 0$ denote the predicted mean precipitation.
The Tweedie distribution belongs to the exponential dispersion family and satisfies \citep{jorgensen1997,dunn2005}
\[
\mathrm{Var}(Y_i)=\phi \mu_i^{p},
\]
where $\phi>0$ is the dispersion parameter and $p$ is the Tweedie variance power.
In this study, the exponent $p$ was treated as a tunable hyperparameter.

The sample weights were defined as a monotone nondecreasing piecewise linear function of the observed precipitation.
Let the knot positions be
\[
x_0=0,\quad x_1=5,\quad x_2=10,\quad x_3=15,\quad x_4=20\ \mathrm{mm},
\]
and the corresponding weight values be
\[
w_0,\ w_1,\ w_2,\ w_3,\ w_4.
\]
A monotonicity constraint,
\[
1 = w_0 \le w_1 \le w_2 \le w_3 \le w_4,
\]
was imposed.
Then, the sample weight function $w(y)$ was defined by linear interpolation between adjacent knots as follows:
\[
w(y)=
\begin{cases}
w_0 + \dfrac{w_1-w_0}{x_1-x_0}(y-x_0), & x_0 \le y \le x_1,\\[8pt]
w_1 + \dfrac{w_2-w_1}{x_2-x_1}(y-x_1), & x_1 < y \le x_2,\\[8pt]
w_2 + \dfrac{w_3-w_2}{x_3-x_2}(y-x_2), & x_2 < y \le x_3,\\[8pt]
w_3 + \dfrac{w_4-w_3}{x_4-x_3}(y-x_3), & x_3 < y.
\end{cases}
\]
In the implementation, the resulting weights were clipped to the range
\[
1 \le w(y) \le w_{\max},
\]
where $w_{\max}=10$.
Thus, samples with higher precipitation intensity contributed more strongly to training, while the monotonicity constraint maintained the stability and interpretability of the weighting scheme.

Using these weights, the precipitation model was trained by minimizing the weighted Tweedie loss
\[
\mathcal{L}_{\mathrm{train}}
=
\sum_{i=1}^{N} w(y_i)\,\ell_{\mathrm{Tw}}(y_i,\mu_i),
\]
where $\ell_{\mathrm{Tw}}(y_i,\mu_i)$ denotes the Tweedie loss for sample $i$.
In this study, we used the Tweedie objective implemented in LightGBM \citep{ke2017lightgbm,lightgbm_params}.
Except for the precipitation loss function and sample-weighting scheme described below, the input data configuration and the tree-related LightGBM settings were the same as those of the ``around all tune'' model.

Although the model was trained using the weighted Tweedie loss, the hyperparameters were selected in Optuna based on an event-based evaluation score, because the purpose of weighting was to improve performance at relevant precipitation thresholds.
The tuned hyperparameters were the Tweedie variance power $p$ and the parameters determining the shape of the weight function, and the tuning was carried out separately for each observation site because the behavior of the weighted Tweedie model showed strong site dependence.
A monotone nondecreasing constraint was imposed on the weight function so that the weight would not decrease as precipitation intensity increased.

Using the event-based metrics defined in Section~\ref{subsubsec:event_metrics_precip}, the TS values across thresholds were aggregated using a geometric mean:
\[
S_{\mathrm{geo}}
=
\exp\!\left(
\frac{1}{|\mathcal{T}|}
\sum_{\tau \in \mathcal{T}}
\log\bigl(\max(\mathrm{TS}(\tau),\varepsilon)\bigr)
\right).
\]

Furthermore, to suppress overprediction or underprediction of event frequency, a penalty was imposed on the forecast Bias at selected thresholds.
Using this, the total score employed in Optuna was defined as
\[
S_{\mathrm{total}}
=
S_{\mathrm{geo}}
-
\sum_{\tau \in \mathcal{T}_{\mathrm{bias}}}
\gamma_{\tau}\, \left| \log\bigl(\max(\mathrm{Bias}(\tau),b_{\min})\bigr) \right|.
\]
Here, $\gamma_{\tau}\ge 0$ denotes the penalty coefficient, $\mathcal{T}_{\mathrm{bias}}$ denotes the set of thresholds to which the bias penalty is applied, and $b_{\min}$ is a small positive constant for numerical stability.
This objective function was intended to encourage models with balanced detection skill across multiple thresholds while reducing substantial overprediction or underprediction of event frequency.

In the present implementation, the thresholds used for optimization were set to
\[
\mathcal{T}=\{1,2,\ldots,20\}\ \mathrm{mm},
\]
and equal weights were used for all thresholds in the geometric mean.
In addition, the bias penalty was applied at 1, 5, 10, and 15 mm with $\gamma_{\tau}=0.05$ for all of these thresholds, and $\varepsilon = 10^{-6}$ and $b_{\min} = 10^{-6}$ were set for numerical stability.
The resulting model is referred to as the weighted Tweedie model in the following sections.

\FloatBarrier

\section{Results and Discussion}
\label{sec:results_and_discussion}

\subsection{Model Comparison}

\subsubsection{RMSE and ME}
First, Table 5 and Table 6 present the RMSE and ME of each model, as
well as for MSM and MSMG, on the validation and test datasets.

For RMSE, both validation and test datasets show that the models using LightGBM outperform MSM and MSMG. Similarly, the neural network and CNN models perform worse compared to those using LightGBM. Moreover, models incorporating data from surrounding grids (around) exhibit slightly better performance than those using only the nearest grid (1grid). This indicates that leveraging features from surrounding grids and reducing them through feature selection allows for the inclusion of informative data that improves accuracy. The correlation analysis-based feature selection (FS0) of surrounding grid data yielded the best results in most cases. Among the three feature selection methods (FS1 - FS3) described in \Cref{subsec:feature_selection}, no performance improvement was observed. For the number of features in FS1 to FS3, we used the values that performed best for each feature selection method and each output variable in \Cref{subsec:comparison_of_feature_selection_methods}.

For ME, models using LightGBM tended to reduce bias in both validation and test datasets. However, for precipitation in the validation dataset, MSM achieved the best ME, while for temperature in the test dataset, MSMG performed the best.

For subsequent predictions in this study, we used the “around all tune” model, which showed the best overall performance; here, “around” indicates the use of surrounding-grid predictors, “all” indicates that all selected predictors were retained without further reduction, and “tune” indicates hyperparameter optimization.

\begin{figure}[!htbp]
    \centering
    \begin{subfigure}{\textwidth}
        \centering
        \includegraphics[width=0.9\textwidth]{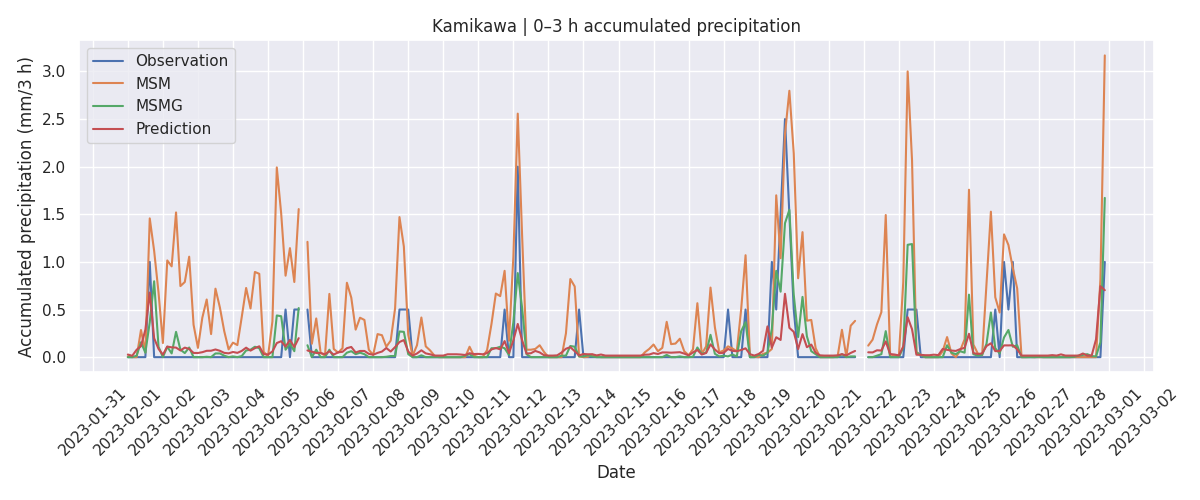}
        \caption{Accumulated precipitation for the 0--3 h forecast interval at Kamikawa}
    \end{subfigure}

    \vspace{0.3em}

    \begin{subfigure}{\textwidth}
        \centering
        \includegraphics[width=0.9\textwidth]{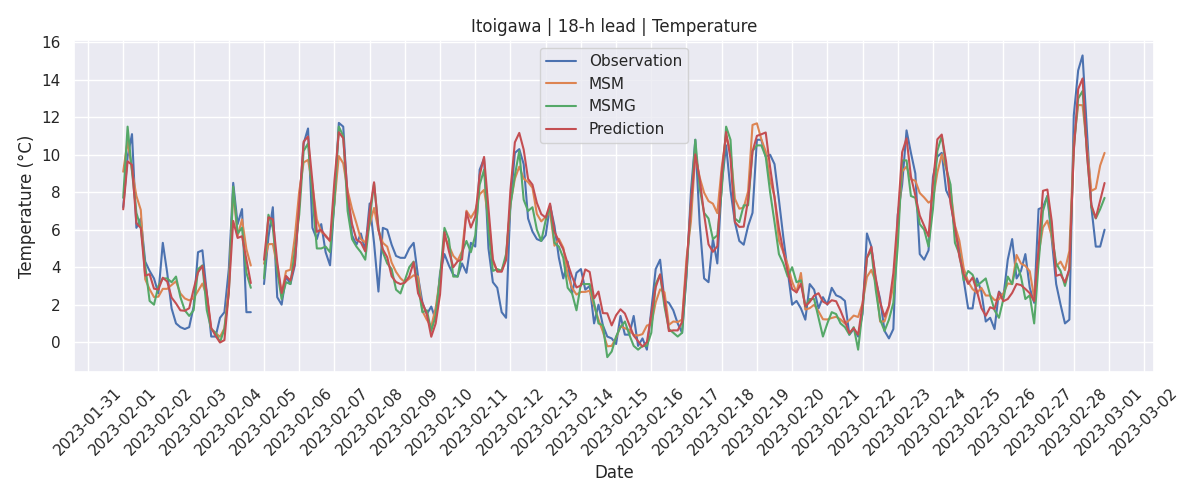}
        \caption{18-h lead temperature at Itoigawa}
    \end{subfigure}

    \vspace{0.3em}

    \begin{subfigure}{\textwidth}
        \centering
        \includegraphics[width=0.9\textwidth]{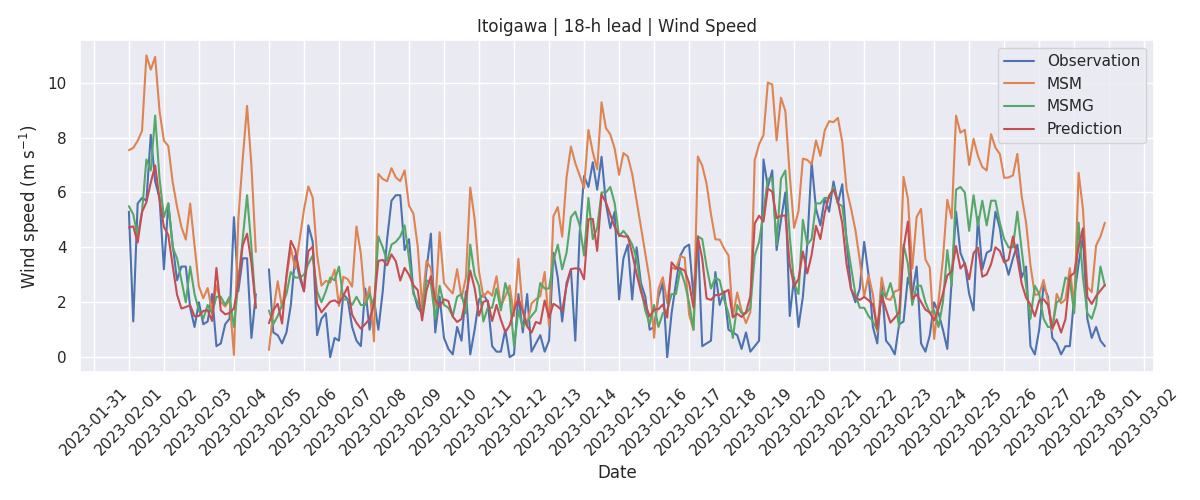}
        \caption{18-h lead wind speed at Itoigawa}
    \end{subfigure}

    \caption{Examples of prediction results for the test data in February, compared with MSM, MSMG, and observed values. Panel (a) shows accumulated precipitation for the 0--3 h forecast interval, while panels (b) and (c) show temperature and wind speed at the 18-h forecast lead time, respectively.}
    \label{fig:line_plot}
\end{figure}

\Cref{fig:line_plot} shows the prediction results for test data compared with MSM, MSMG, and observed values at selected locations and prediction times. In the case of 3-hour precipitation at Kamikawa and 18-hour wind speed at Itoigawa, post-processing models corrected the apparent large bias. On the other hand, for 18-hour temperature at Itoigawa, partial corrections by the post-processing model were observed.

\begin{table}[!htbp]
\centering
\caption{RMSE and ME for validation data. ``1grid'' represents models using data from only one grid, ``around'' indicates the use of surrounding grid data, ``nn'' denotes neural network, ``all'' refers to data before three feature selection methods (FS1 - FS3), ``pcc'' indicates the correlation coefficient-based method (FS1), ``mi'' denotes the mutual information method (FS2), ``gbm'' represents the feature importance method by LightGBM (FS3), and ``tune'' refers to parameter tuning. Averages are calculated across prediction times and locations.}
\label{tab:val_overall}
\begin{tabular}{l|rrr|rrr}
\toprule
 & \multicolumn{3}{c}{RMSE} & \multicolumn{3}{c}{ME} \\
 & Precip. & Temp. & Wind & Precip. & Temp. & Wind \\
\midrule
MSM & 2.948 & 1.836 & 2.096 & 0.048 & -0.405 & 0.753 \\
1grid nn & 2.555 & 1.472 & 1.167 & -0.075 & 0.130 & -0.013 \\
1grid all & 2.510 & 1.364 & 1.070 & -0.007 & 0.084 & -0.018 \\
1grid pcc & 2.526 & 1.367 & 1.070 & -0.003 & 0.085 & -0.018 \\
1grid mi & 2.535 & 1.420 & 1.070 & \textbf{-0.001} & 0.096 & \textbf{-0.016} \\
1grid gbm & 2.518 & 1.363 & 1.073 & -0.006 & 0.085 & -0.018 \\
around all & 2.501 & 1.332 & 1.050 & \textbf{0.001} & 0.081 & -0.025 \\
around pcc & 2.496 & 1.344 & 1.053 & 0.006 & 0.082 & -0.026 \\
around mi & 2.505 & 1.342 & 1.051 & 0.004 & 0.083 & -0.026 \\
around gbm & 2.502 & 1.333 & 1.051 & -0.002 & 0.080 & -0.027 \\
around all tune & \textbf{2.489} & \textbf{1.326} & \textbf{1.046} & 0.009 & \textbf{0.074} & -0.023 \\
CNN &  & 1.487 &  &  & 0.086 &  \\
\bottomrule
\end{tabular}

\end{table}

\begin{table}[!htbp]
\centering
\caption{RMSE and ME for test data.}
\label{tab:test_overall}
\begin{tabular}{l|rrr|rrr}
\toprule
 & \multicolumn{3}{c}{RMSE} & \multicolumn{3}{c}{ME} \\
 & Precip. & Temp. & Wind & Precip. & Temp. & Wind \\
\midrule
MSM & 2.759 & 1.811 & 1.999 & 0.085 & -0.340 & 0.673 \\
MSMG & 2.726 & 1.394 & 1.237 & 0.107 & \textbf{0.014} & 0.159 \\
1grid nn & 2.439 & 1.485 & 1.175 & -0.064 & 0.143 & -0.060 \\
1grid all & 2.387 & 1.368 & 1.075 & 0.020 & 0.110 & -0.058 \\
1grid pcc & 2.390 & 1.369 & 1.076 & \textbf{0.016} & 0.109 & -0.058 \\
1grid mi & 2.409 & 1.428 & 1.075 & 0.028 & 0.117 & \textbf{-0.057} \\
1grid gbm & 2.392 & 1.368 & 1.078 & 0.017 & 0.111 & -0.058 \\
around all & 2.358 & 1.341 & 1.053 & 0.031 & 0.111 & -0.062 \\
around pcc & 2.374 & 1.349 & 1.056 & 0.030 & 0.124 & -0.062 \\
around mi & 2.381 & 1.347 & 1.055 & 0.033 & 0.119 & -0.061 \\
around gbm & 2.361 & 1.341 & \textbf{1.052} & 0.028 & 0.112 & -0.063 \\
around all tune & \textbf{2.344} & \textbf{1.335} & \textbf{1.052} & 0.033 & 0.106 & -0.059 \\
CNN &  & 1.501 &  &  & 0.179 &  \\
\bottomrule
\end{tabular}
\end{table}

\subsubsection{Statistical Tests and Event-Based Metrics}

For temperature and wind speed, Table 7 summarizes the paired \(t\)-test and
Wilcoxon signed-rank test results comparing the main model (around all
tune) with the baseline models (MSM and MSMG), based on the metrics
defined in Section~\ref{subsubsec:statistical_tests}. The table reports
\(\mathrm{MAE}_{\mathrm{main}}\), \(\mathrm{MAE}_{\mathrm{base}}\), the standard deviation of the paired
differences, and the corresponding one-sided \(p\)-values. Because of the very
large sample size, these test results should be interpreted only as
supplementary information, and practical significance should be assessed
mainly from the error metrics themselves.

For precipitation, Table 8 summarizes the event-based metrics defined in
Section~\ref{subsubsec:event_metrics_precip}---\(\mathrm{TS}\),
\(\mathrm{POD}\), \(\mathrm{FAR}\), and \(\mathrm{Bias}\)---for four
precipitation thresholds. Here, ``around all tune'' denotes our original
LightGBM-based post-processing model, and ``weighted Tweedie'' denotes the weighted
Tweedie model.

Overall, MSMG showed favorable performance among the compared models.
The ``around all tune'' model achieved the highest POD at the 1.0 mm threshold, indicating high sensitivity to light precipitation events.
However, its FAR and Bias were also the largest, suggesting that it tended to overpredict precipitation occurrence.
At thresholds of 5.0 mm and above, its TS and POD became lower than those of MSM and MSMG, and its Bias fell well below 1 at 10.0 mm and 15.0 mm, indicating underdetection of heavier precipitation events.
Compared with this behavior, the weighted Tweedie model improved TS and Bias at 10.0 mm and increased POD at 5.0 mm.
Although its overall performance was still slightly below that of MSMG, it showed a better balance across the precipitation thresholds.

\begin{table}[!htbp]
\centering
\caption{
Paired $t$-test and Wilcoxon signed-rank test results comparing the LightGBM ``around all tune'' model (main model) with the baseline models for temperature and wind speed.
The table reports the mean absolute errors (MAEs) of the main and baseline models, the standard deviation of the paired differences in absolute error ($\mathrm{SD_{diff}}$), and the corresponding one-sided $p$-values.
All tests were performed over $N=682{,}305$ paired samples.}
\label{tab:paired_tests_overall}

\begin{tabular}{llrrrrr}
\toprule
Variable & Baseline & MAE$_\text{main}$ & MAE$_\text{base}$ & SD$_\text{diff}$ & $p_t$ & $p_W$ \\
\midrule
Temperature & MSM  & 0.9888 & 1.3964 & 1.0840 & $0.000$ & $0.000$ \\
Temperature & MSMG & 0.9888 & 1.0275 & 0.7675 & $0.000$ & $0.000$ \\
Wind Speed  & MSM  & 0.7549 & 1.4186 & 1.3530 & $0.000$ & $0.000$ \\
Wind Speed  & MSMG & 0.7549 & 0.9035 & 0.6980 & $0.000$ & $0.000$ \\
\bottomrule
\end{tabular}
\end{table}

\begin{table}[!htbp]
\centering
\caption{
Event-based verification metrics for precipitation, pooled over all locations and forecast lead times.
Higher TS and POD indicate better detection skill, lower FAR indicates fewer false alarms, and Bias values closer to 1 indicate better frequency matching.
Here, ``around all tune'' denotes the original LightGBM-based post-processing model.
}
\label{tab:precip_event_scores_overall}
\setlength{\tabcolsep}{5pt}
\begin{tabular}{ccrrrr}
\toprule
Threshold (mm) & Model & TS & POD & FAR & Bias \\
\midrule
1.0  & MSM       & 0.4311 & 0.6373 & 0.4287 & 1.1156 \\
1.0  & MSMG      & \textbf{0.4555} & 0.6558 & \textbf{0.4014} & \textbf{1.0956} \\
1.0  & around all tune      & 0.4293 & 0.6931 & 0.4699 & 1.3075 \\
1.0  & weighted Tweedie & 0.4435 & \textbf{0.7005} & 0.4527 & 1.2800 \\
\midrule
5.0  & MSM       & 0.3226 & 0.5061 & 0.5293 & 1.0753 \\
5.0  & MSMG      & \textbf{0.3344} & 0.5014 & 0.4991 & \textbf{1.0009} \\
5.0  & around all tune      & 0.3059 & 0.3943 & \textbf{0.4228} & 0.6831 \\
5.0  & weighted Tweedie & 0.3316 & \textbf{0.5208} & 0.5228 & 1.0912 \\
\midrule
10.0 & MSM       & 0.2450 & 0.4084 & 0.6201 & 1.0749 \\
10.0 & MSMG      & 0.2485 & \textbf{0.4122} & 0.6152 & 1.0713 \\
10.0 & around all tune      & 0.1966 & 0.2356 & \textbf{0.4571} & 0.4340 \\
10.0 & weighted Tweedie & \textbf{0.2520} & 0.3979 & 0.5927 & \textbf{0.9771} \\
\midrule
15.0 & MSM       & 0.1902 & 0.3189 & 0.6796 & \textbf{0.9956} \\
15.0 & MSMG      & \textbf{0.2096} & \textbf{0.3551} & 0.6616 & 1.0492 \\
15.0 & around all tune      & 0.1251 & 0.1420 & \textbf{0.4877} & 0.2772 \\
15.0 & weighted Tweedie & 0.1883 & 0.2719 & 0.6201 & 0.7158 \\
\bottomrule
\end{tabular}
\end{table}

\subsection{Comparison of Feature Selection Methods}
\label{subsec:comparison_of_feature_selection_methods}
\begin{figure}[!htbp]
    \centering
    \begin{subfigure}{\textwidth}
        \centering
        \includegraphics[width=0.62\textwidth]{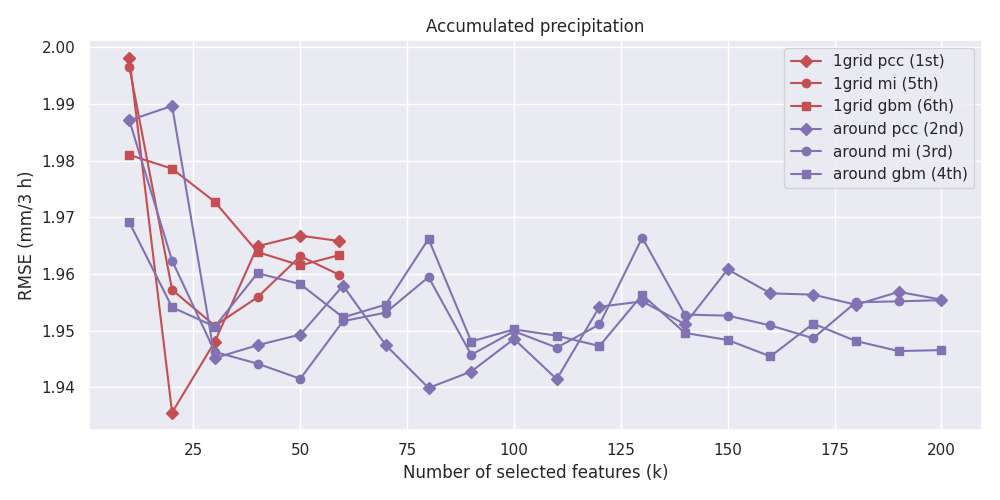}
        \caption{Precipitation}
    \end{subfigure}
    \vspace{0.4em}

    \begin{subfigure}{\textwidth}
        \centering
        \includegraphics[width=0.62\textwidth]{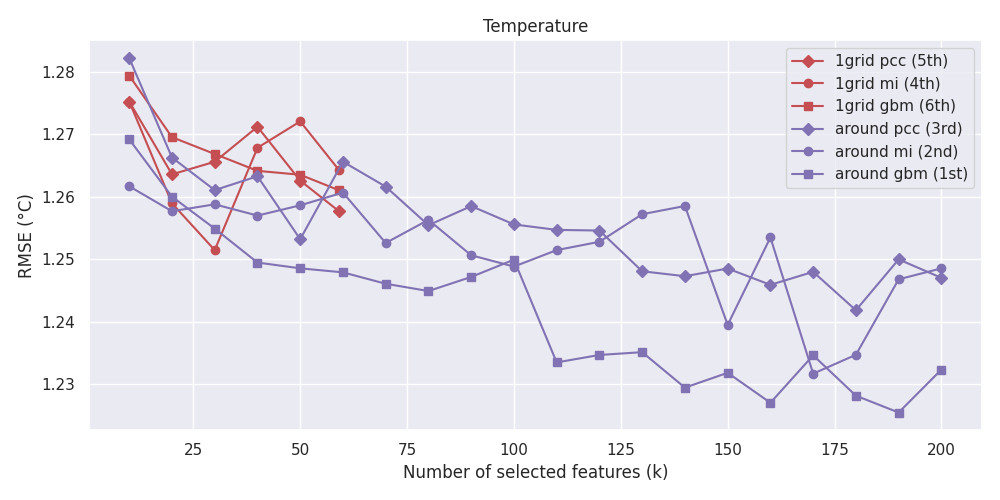}
        \caption{Temperature}
    \end{subfigure}
    \vspace{0.4em}

    \begin{subfigure}{\textwidth}
        \centering
        \includegraphics[width=0.62\textwidth]{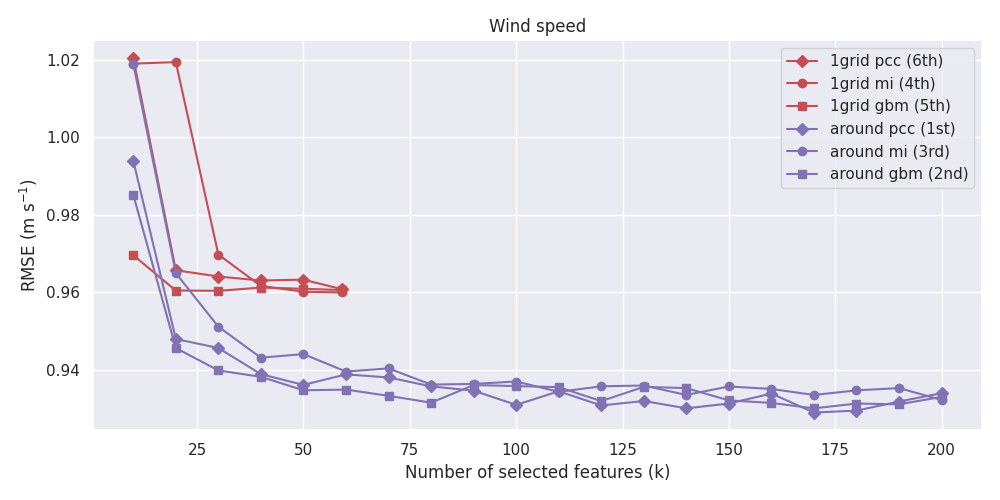}
        \caption{Wind Speed}
    \end{subfigure}
    
    \caption{Changes in accuracy based on the number of selected features for the feature selection methods on validation data in Sekigahara. Results are averaged over prediction times.}
    \label{fig:selection_plot}
\end{figure}

Here, we conducted experiments at a single location in Sekigahara, similar to parameter tuning, to evaluate the influence of the number of selected features in the three feature selection methods (FS1 - FS3) proposed in \Cref{subsec:feature_selection}. For both cases where only the data of the grid closest to the prediction point (1grid) and the data including surrounding grids (around) were used, the three feature selection methods were applied, and models were constructed by incrementally increasing the number of selected features $k$ by 10. \Cref{fig:selection_plot} summarizes the RMSE on the validation data for each feature count $k$. The results indicate that the accuracy tends to improve as the number of features increases, but stabilizes when the number of features exceeds a certain threshold. This suggests that further feature selection after correlation analysis does not enhance accuracy.

\subsection{Results by Location and Prediction Time}
\label{subsec:results_by_location_and_prediction_time}

\subsubsection{Location-Based Results}
\begin{figure}[!htbp]
    \centering
    \begin{subfigure}{\textwidth}
        \centering
        \includegraphics[width=0.80\textwidth]{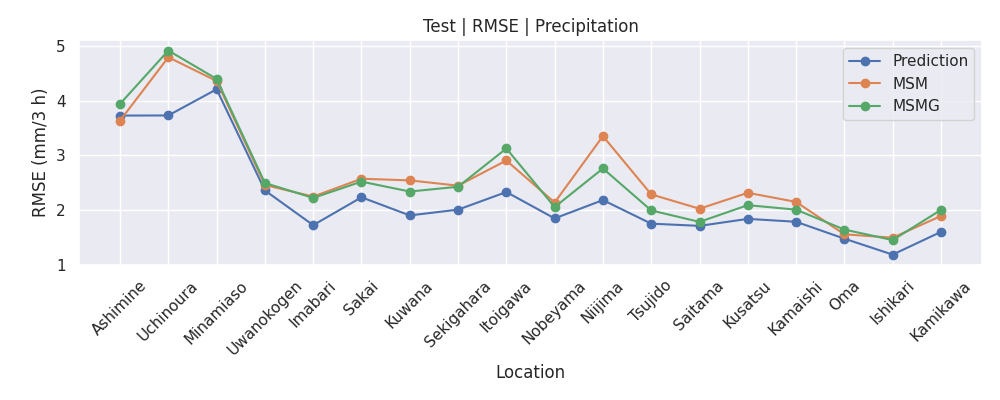}
        \caption{Precipitation}
    \end{subfigure}
    \vspace{0.4em}

    \begin{subfigure}{\textwidth}
        \centering
        \includegraphics[width=0.80\textwidth]{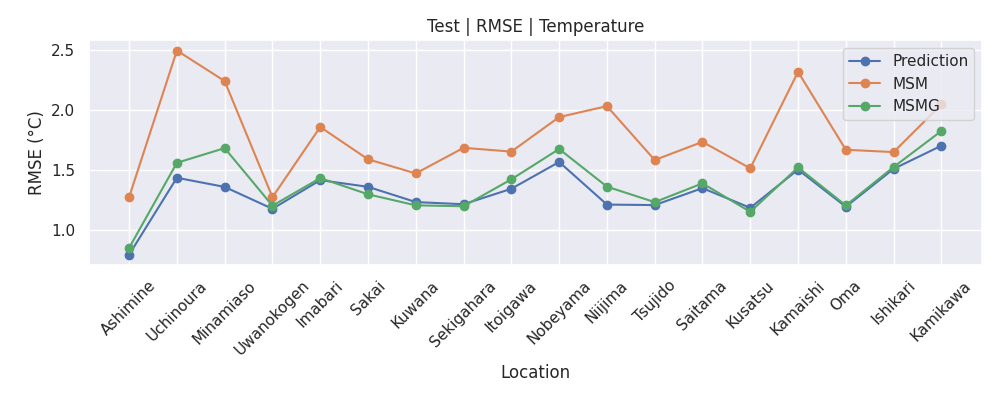}
        \caption{Temperature}
    \end{subfigure}
    \vspace{0.4em}

    \begin{subfigure}{\textwidth}
        \centering
        \includegraphics[width=0.80\textwidth]{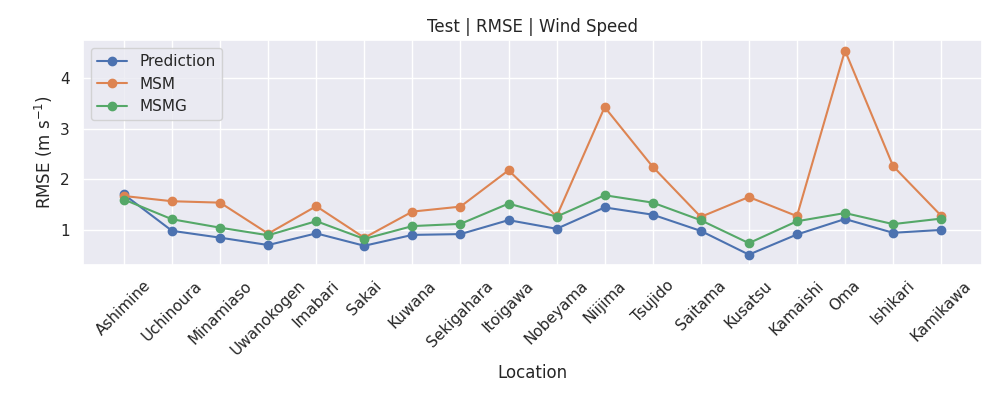}
        \caption{Wind Speed}
    \end{subfigure}
    
    \caption{RMSE by location on test data. Results are averaged over prediction times.}
    \label{fig:pos_rmse}
\end{figure}

\Cref{fig:pos_rmse} summarizes the RMSE on test data for each observation location.
Our post-processing model consistently outperformed MSM and showed comparable or slightly better performance than MSMG across most sites.
For precipitation, relatively large RMSE values were observed in the southwestern part of Japan, including Ashimine and Uchinoura.
This is likely associated with the frequent occurrence of typhoons and strong convective rainfall events in these regions.
In contrast, the lowest precipitation RMSE values were found in inland and mountainous areas such as Kusatsu and Nobeyama.

For wind speed, extremely high RMSE values in MSM were observed in island and coastal areas such as Niijima and Oma.
Our model significantly reduced these errors, achieving RMSE values comparable to those in other regions.
Overall, the regional pattern of RMSE reduction indicates that the post-processing model effectively mitigates the biases and noise that are spatially correlated with topography and surrounding meteorological conditions.

\begin{table}[!htbp]
  \centering
  \caption{Average RMSE by terrain category and model.
  The three terrain categories are defined as follows:
  \textbf{mountain} (Kusatsu, Kamikawa, Nobeyama, Uwanokogen, Minamiaso),
  \textbf{island} (Ashimine, Niijima),
  and \textbf{coastal/plain}
  (Ishikari, Oma, Imabari, Kamaishi, Tsujido, Itoigawa, Uchinoura,
   Sekigahara, Saitama, Kuwana, Sakai).}
  \label{tab:rmse_by_terrain}
  \begin{tabular}{lccccccccc}
    \toprule
    \multirow{2}{*}{Region} &
    \multicolumn{3}{c}{around\_all\_tune} &
    \multicolumn{3}{c}{MSM} &
    \multicolumn{3}{c}{MSMG} \\
    \cmidrule(lr){2-4}\cmidrule(lr){5-7}\cmidrule(lr){8-10}
     & Precip. & Temp. & Wind & Precip. & Temp. & Wind & Precip. & Temp. & Wind \\
    \midrule
    coastal/plain & 1.982 & 1.343 & 1.002 & 2.454 & 1.793 & 1.864 & 2.400 & 1.363 & 1.211 \\
    island        & 2.951 & 1.003 & 1.576 & 3.492 & 1.654 & 2.550 & 3.349 & 1.106 & 1.647 \\
    mountain      & 2.369 & 1.398 & 0.822 & 2.630 & 1.804 & 1.337 & 2.605 & 1.508 & 1.038 \\
    \bottomrule
  \end{tabular}
\end{table}

As shown in Table~\ref{tab:rmse_by_terrain}, the selected \textit{around\_all\_tune} model achieved the lowest RMSE across all terrain types and variables.
The improvement was most pronounced in mountainous and island areas, where spatial heterogeneity in meteorological conditions is large.
This demonstrates that incorporating spatial context from surrounding grids improves model robustness under complex terrain and coastal influences.

\subsubsection{Prediction-Time-Based Results}
\begin{figure}[!htbp]
    \centering
    \begin{subfigure}{\textwidth}
        \centering
        \includegraphics[width=0.80\textwidth]{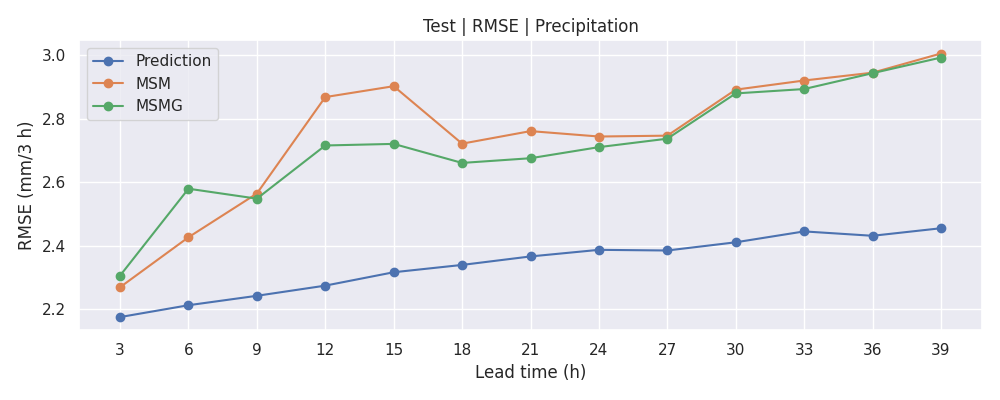}
        \caption{Precipitation}
    \end{subfigure}
    \vspace{0.4em}

    \begin{subfigure}{\textwidth}
        \centering
        \includegraphics[width=0.80\textwidth]{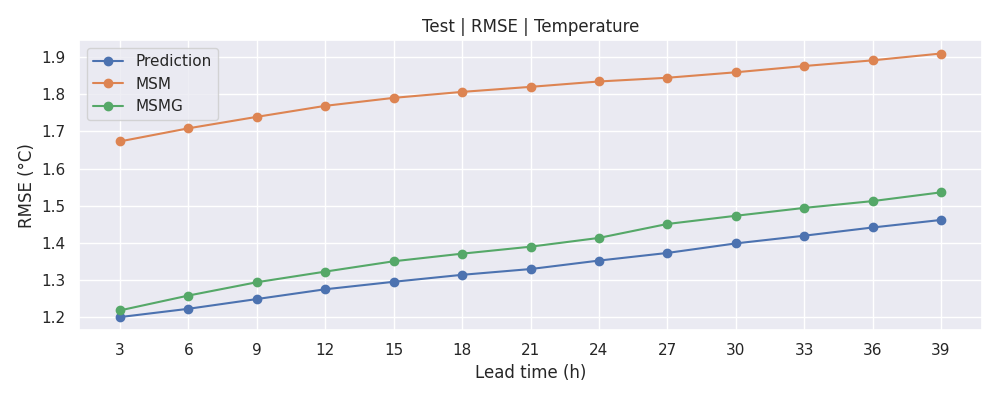}
        \caption{Temperature}
    \end{subfigure}
    \vspace{0.4em}

    \begin{subfigure}{\textwidth}
        \centering
        \includegraphics[width=0.80\textwidth]{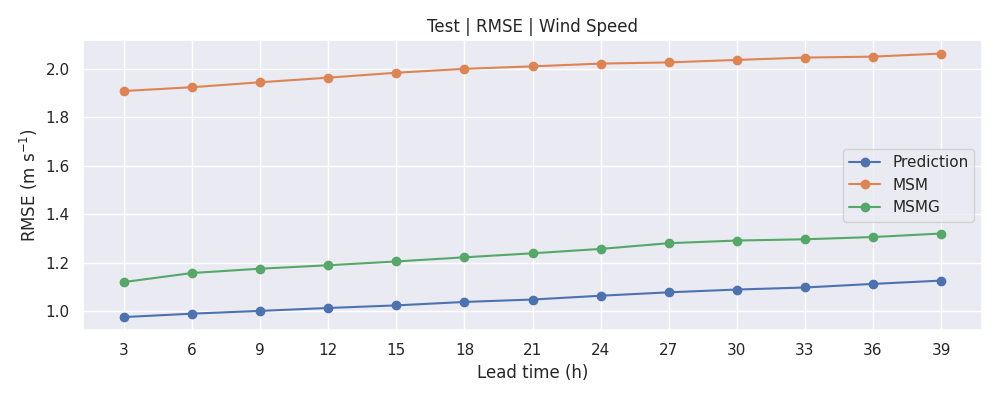}
        \caption{Wind Speed}
    \end{subfigure}
    
    \caption{RMSE by prediction time on test data. Results are averaged over locations.}
    \label{fig:time_rmse}
\end{figure}

\Cref{fig:time_rmse} summarizes the RMSE on test data for each prediction time.
In terms of RMSE, the selected around\_all\_tune model outperformed both MSM and MSMG at almost all forecast lead times.
For all methods, RMSE increased as prediction time became longer,
which is a common characteristic of numerical weather prediction (NWP) models, where forecast accuracy declines with lead time.
A similar trend was observed in our post-processing results,
indicating that the residual errors of the base NWP propagate into the corrected predictions as forecast uncertainty accumulates.
Nevertheless, the selected around\_all\_tune model maintained lower RMSE than MSM and, in many cases, MSMG across forecast horizons, confirming its stability over time.

\subsubsection{Scatter Plots Against Observations}
\begin{figure}[!htbp]
    \centering
    \begin{subfigure}{0.32\textwidth}
        \centering
        \includegraphics[width=0.82\textwidth]{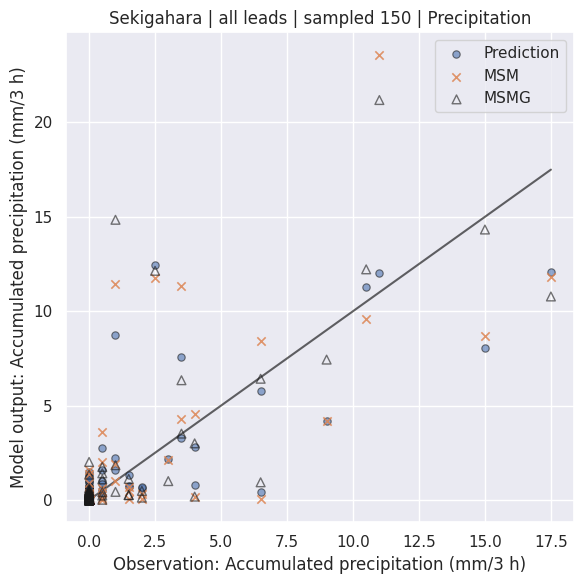}
        \caption{Precipitation}
    \end{subfigure}\hfill
    \begin{subfigure}{0.32\textwidth}
        \centering
        \includegraphics[width=0.82\textwidth]{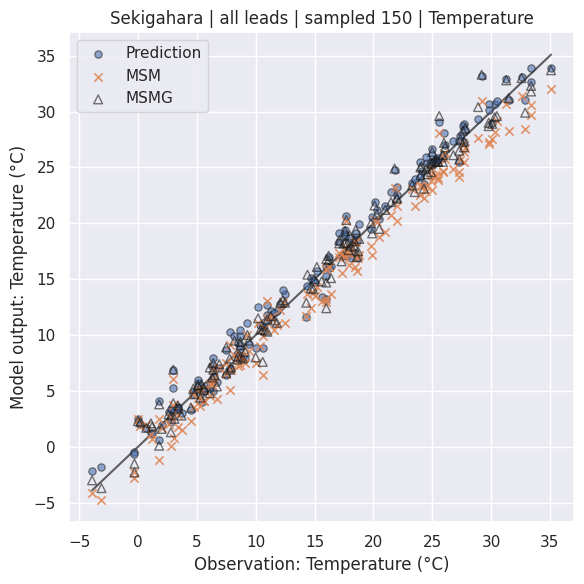}
        \caption{Temperature}
    \end{subfigure}\hfill
    \begin{subfigure}{0.32\textwidth}
        \centering
        \includegraphics[width=0.82\textwidth]{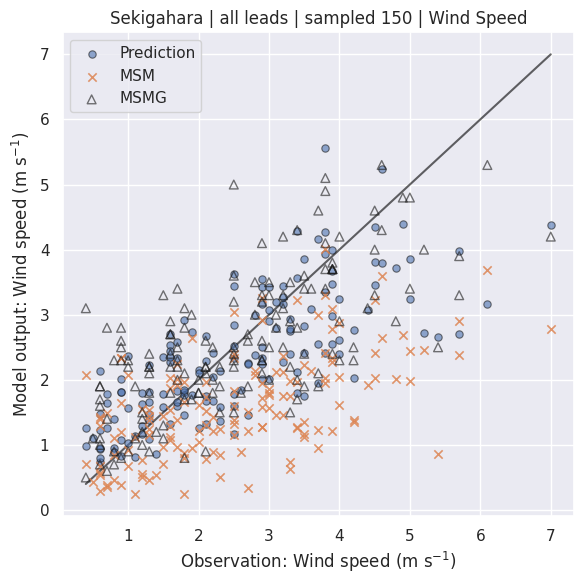}
        \caption{Wind speed}
    \end{subfigure}

    \caption{Scatter plots at Sekigahara on the test period (all lead times pooled). Each panel shows observations versus model outputs for the same randomly sampled points.}
    \label{fig:scatter_sekigahara}
\end{figure}

Figure~\ref{fig:scatter_sekigahara} compares observations with the model outputs at Sekigahara during the test period.
For readability, all lead times are pooled and a random subset of 150 samples is plotted.

\subsection{Site-Dependent Behavior of the Weighted Tweedie Model}
\label{subsec:site_dependent_behavior}

The effect of the weighted Tweedie model was strongly site dependent.
To illustrate this point, we compare two representative examples: Kamikawa, where the improvement was limited, and Saitama at the 18-h lead time, where the weighted Tweedie model showed clearer improvement in the time-series comparison.

\begin{figure}[!htbp]
    \centering
    \begin{subfigure}{\textwidth}
        \centering
        \includegraphics[width=0.9\textwidth]{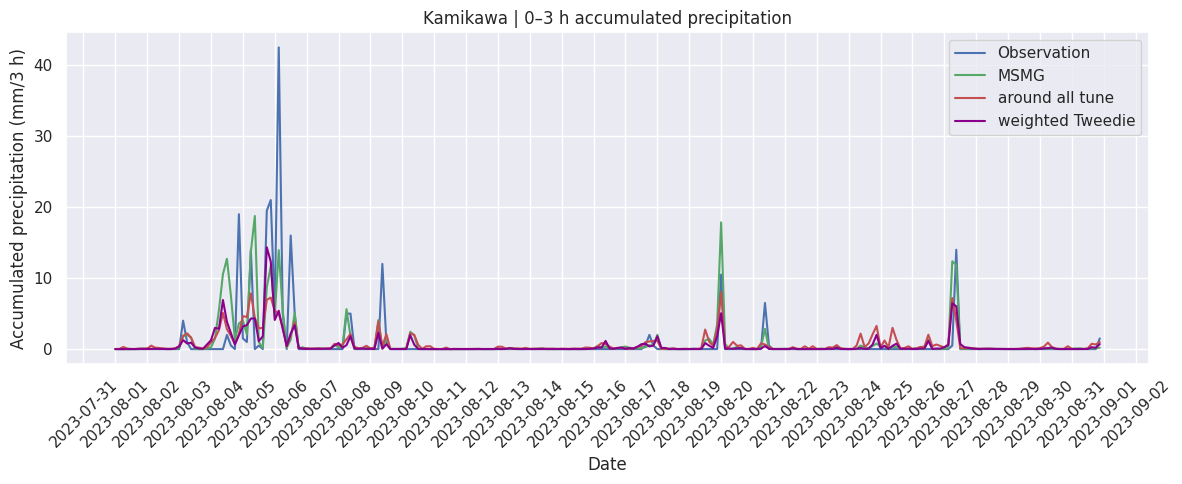}
        \caption{Accumulated precipitation for the 0--3 h forecast interval at Kamikawa}
    \end{subfigure}
    \vspace{0.4em}

    \begin{subfigure}{\textwidth}
        \centering
        \includegraphics[width=0.9\textwidth]{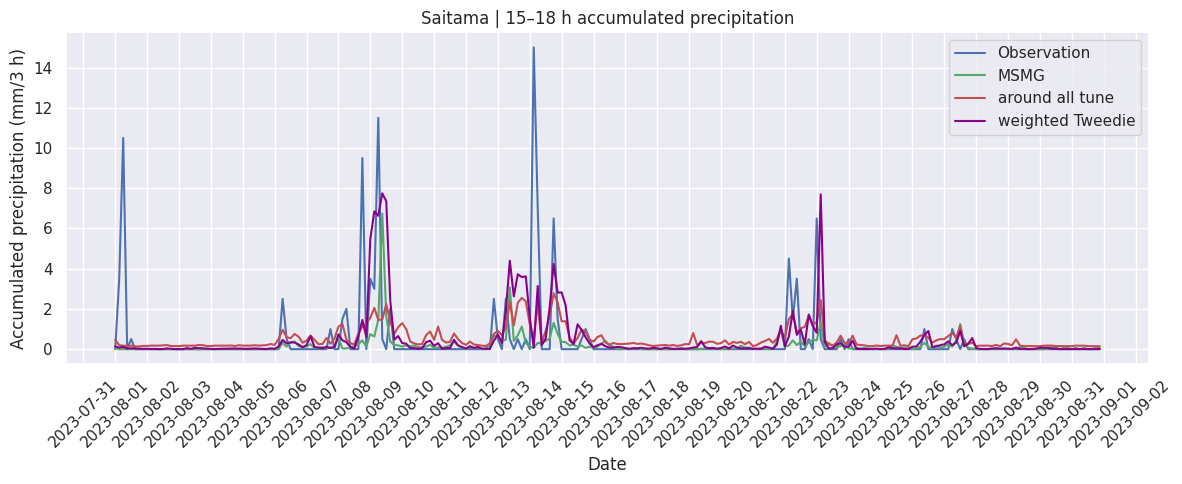}
        \caption{Accumulated precipitation for the 15--18 h forecast interval at Saitama}
    \end{subfigure}
    
\caption{Comparison of precipitation forecasts for two representative cases. Panel (a) shows Kamikawa, where the weighted Tweedie model remained close to the original ``around all tune'' model and the improvement was limited. Panel (b) shows Saitama, where the weighted Tweedie model showed improved event-oriented behavior for several precipitation events.}
    \label{fig:line_improved}
\end{figure}

Figure~\ref{fig:line_improved}(a) shows the 0--3 h accumulated precipitation at Kamikawa.
In this case, the weighted Tweedie model remained close to the original ``around all tune'' model, and the difference between the two predictions was small for many events.
This behavior is consistent with the site-wise aggregated verification results summarized in Table~\ref{tab:site_case_precip_metrics}.
At Kamikawa, the weighted Tweedie model produced only a slight change at the 1.0 mm threshold and did not improve TS at 5.0 mm, although TS and POD were slightly improved at 10.0 and 15.0 mm.
Overall, however, the weighted Tweedie model still remained clearly below MSMG for this site, indicating that the modification was not sufficient to overcome the difficulty of precipitation forecasting at this site.

By contrast, Figure~\ref{fig:line_improved}(b) shows a case at Saitama for the 15--18 h accumulated precipitation, where the weighted Tweedie model more clearly modified the predicted peaks and, for several events, tracked the observed precipitation more closely than ``around all tune''.
This behavior is also reflected in the site-wise aggregated results in Table~\ref{tab:site_case_precip_metrics}.
At Saitama, the weighted Tweedie model improved TS relative to ``around all tune'' at all four thresholds and substantially increased POD at 5.0, 10.0, and 15.0 mm, while also moving Bias much closer to 1 than the original model at 5.0 and 10.0 mm.
At 15.0 mm, the weighted Tweedie model still showed the best TS among the compared models, although its Bias remained below 1.
These results suggest that the weighted Tweedie model can improve event-oriented behavior for some sites, especially when the original model underdetects moderate to heavy precipitation.

Taken together, these examples indicate that the weighted Tweedie model is useful mainly as a site-dependent adjustment rather than as a uniformly superior replacement for the original LightGBM model.
This observation is consistent with the need for site-wise hyperparameter tuning discussed in Section~\ref{subsec:weighted_tweedie}.

\begin{table}[!htbp]
\centering
\caption{
Site-wise precipitation verification metrics for two representative cases.
Higher TS and POD indicate better detection skill, while Bias values closer to 1 indicate better frequency matching.
}
\label{tab:site_case_precip_metrics}
\setlength{\tabcolsep}{4pt}
\begin{tabular}{llcrrr}
\toprule
Site & Model & Threshold (mm) & TS & POD & Bias \\
\midrule
\multirow{12}{*}{Kamikawa}
& MSMG              & 1.0  & 0.3802 & \textbf{0.5501} & \textbf{0.9933} \\
& around all tune   & 1.0  & 0.3817 & 0.5311 & 0.9367 \\
& weighted Tweedie  & 1.0  & \textbf{0.3863} & 0.5319 & 0.9276 \\
& MSMG              & 5.0  & \textbf{0.2447} & \textbf{0.3897} & \textbf{0.9464} \\
& around all tune   & 5.0  & 0.1541 & 0.1852 & 0.3492 \\
& weighted Tweedie  & 5.0  & 0.1264 & 0.1488 & 0.3057 \\
& MSMG              & 10.0 & \textbf{0.1278} & \textbf{0.2802} & \textbf{0.9973} \\
& around all tune   & 10.0 & 0.0267 & 0.0275 & 0.0687 \\
& weighted Tweedie  & 10.0 & 0.0471 & 0.0522 & 0.0962 \\
& MSMG              & 15.0 & \textbf{0.0667} & \textbf{0.1703} & \textbf{0.9725} \\
& around all tune   & 15.0 & 0.0000 & 0.0000 & 0.0000 \\
& weighted Tweedie  & 15.0 & 0.0055 & 0.0055 & 0.0055 \\
\midrule
\multirow{12}{*}{Saitama}
& MSMG              & 1.0  & \textbf{0.4391} & 0.6383 & \textbf{1.0919} \\
& around all tune   & 1.0  & 0.4150 & 0.6911 & 1.3563 \\
& weighted Tweedie  & 1.0  & 0.4330 & \textbf{0.7160} & 1.3696 \\
& MSMG              & 5.0  & \textbf{0.3549} & \textbf{0.4994} & \textbf{0.9065} \\
& around all tune   & 5.0  & 0.2308 & 0.2521 & 0.3444 \\
& weighted Tweedie  & 5.0  & 0.3429 & 0.4959 & 0.9420 \\
& MSMG              & 10.0 & 0.1935 & 0.3231 & 0.9923 \\
& around all tune   & 10.0 & 0.1493 & 0.1538 & 0.1846 \\
& weighted Tweedie  & 10.0 & \textbf{0.3395} & \textbf{0.4962} & \textbf{0.9577} \\
& MSMG              & 15.0 & 0.2048 & \textbf{0.3675} & 1.1624 \\
& around all tune   & 15.0 & 0.0678 & 0.0684 & 0.0769 \\
& weighted Tweedie  & 15.0 & \textbf{0.2600} & 0.3333 & \textbf{0.6154} \\
\bottomrule
\end{tabular}
\end{table}

\FloatBarrier

\section{Conclusion}
\label{sec:conclusion}
In this study, we developed machine learning-based post-processing models for precipitation, temperature, and wind speed using observation data and MSM data across 18 locations in Japan, including plains, mountainous areas, and islands. The main findings are summarized as follows:

\begin{enumerate}
\item The use of surrounding grid data together with correlation-based feature selection improved prediction accuracy compared with using only the single nearest grid. This result supports the effectiveness of incorporating spatial information from surrounding grids while controlling feature dimensionality through feature selection. In contrast, further feature selection beyond this correlation-based reduction did not provide additional improvement in our experiments.

\item In our experimental setting, the LightGBM-based models achieved lower RMSE than the tested neural-network baselines, including the reproduced CNN baseline, and also generally achieved lower RMSE than the JMA guidance product MSMG.
However, because the neural-network comparison was limited to specific baseline implementations, the present results do not exclude the possibility that stronger deep-learning architectures could perform better.

\item For precipitation, the results showed that optimizing only standard regression-oriented objectives is not necessarily sufficient, especially when the forecasting goal is to improve the prediction of moderate and heavy rainfall events. To address this issue, we introduced Tweedie-based loss functions and event-weighted training strategies. These approaches improved event-oriented behavior, including better detection of stronger rainfall events at some sites and thresholds. However, the gains were not uniform across locations and conditions, indicating that further improvement is still needed. This site dependence suggests that inter-station differences related to topography and site characteristics, such as latitude, longitude, and altitude, may need to be considered more explicitly in future work.

\end{enumerate}

To complement the representative examples and pooled results shown in the main text, detailed verification results by site and forecast lead time are provided in the Supplementary Material. These include RMSE and ME of MSM, MSMG, and ``around all tune'' for precipitation, temperature, and wind speed, as well as precipitation event-based metrics (TS, POD, FAR, and Bias), including those of the weighted Tweedie model.


\bigskip

\noindent \textbf{Data and Code Availability}\\
The code used in this study is available at: \url{https://github.com/ttakenawa/PostMSM}.
Due to data licensing restrictions, the repository provides a synthetic/derived dataset for demonstration, which may not exactly reproduce the paper results.

\bigskip

\noindent \textbf{Declaration of generative AI and AI-assisted technologies in the manuscript preparation process}\\
During the preparation of this manuscript, the authors used ChatGPT to assist with code editing and language proofreading. All AI-assisted outputs were reviewed and revised by the authors as needed, and the authors take full responsibility for the content of the manuscript.

\bigskip

\noindent \textbf{Acknowledgements}\\
The authors thank the reviewers for their careful reading and helpful comments. The first author was supported by the WISE Program for the Development of AI Professionals in the Marine Industry at Tokyo University of Marine Science and Technology; the second author was supported by the Japan Society for the Promotion of Science, Grant-in-Aid (C) (22K03383).

\FloatBarrier

\begin{thebibliography}{28}
\providecommand{\natexlab}[1]{#1}
\providecommand{\url}[1]{\texttt{#1}}
\expandafter\ifx\csname urlstyle\endcsname\relax
  \providecommand{\doi}[1]{doi: #1}\else
  \providecommand{\doi}{doi: \begingroup \urlstyle{rm}\Url}\fi

\bibitem[Akiba et~al.(2019)Akiba, Sano, Yanase, Ohta, and
  Koyama]{akiba2019optuna}
Takuya Akiba, Shotaro Sano, Toshihiko Yanase, Takeru Ohta, and Masanori Koyama.
\newblock Optuna: A next-generation hyperparameter optimization framework.
\newblock In \emph{Proceedings of the 25th ACM SIGKDD international conference
  on knowledge discovery \& data mining}, pages 2623--2631, 2019.

\bibitem[Bauer et~al.(2015)Bauer, Thorpe, and Brunet]{bauer2015quiet}
Peter Bauer, Alan Thorpe, and Gilbert Brunet.
\newblock The quiet revolution of numerical weather prediction.
\newblock \emph{Nature}, 525\penalty0 (7567):\penalty0 47--55, 2015.

\bibitem[Bedi and Toshniwal(2018)]{bedi2018attribute}
Jatin Bedi and Durga Toshniwal.
\newblock Attribute selection based on correlation analysis.
\newblock In \emph{Advances in big data and cloud computing}, pages 51--61.
  Springer, 2018.

\bibitem[Chandrashekar and Sahin(2014)]{chandrashekar2014survey}
Girish Chandrashekar and Ferat Sahin.
\newblock A survey on feature selection methods.
\newblock \emph{Computers \& electrical engineering}, 40\penalty0 (1):\penalty0
  16--28, 2014.

\bibitem[Dunn and Smyth(2005)]{dunn2005}
Peter~K. Dunn and Gordon~K. Smyth.
\newblock Series evaluation of tweedie exponential dispersion model densities.
\newblock \emph{Statistics and Computing}, 15\penalty0 (4):\penalty0 267--280,
  2005.

\bibitem[Grinsztajn et~al.(2022)Grinsztajn, Oyallon, and
  Varoquaux]{grinsztajn2022tree}
L{\'e}o Grinsztajn, Edouard Oyallon, and Ga{\"e}l Varoquaux.
\newblock Why do tree-based models still outperform deep learning on typical
  tabular data?
\newblock \emph{Advances in neural information processing systems},
  35:\penalty0 507--520, 2022.

\bibitem[Hieta and Partio(2025)]{hieta2025operational}
Leila Hieta and Mikko Partio.
\newblock Operational machine learning post-processing of short-range
  temperature, humidity, wind speed and gust forecasts.
\newblock \emph{Meteorological Applications}, 32:\penalty0 e70074, 2025.

\bibitem[Iwase and Takenawa(2024)]{iwase2024interpolation}
Kazuma Iwase and Tomoyuki Takenawa.
\newblock Interpolation of mountain weather forecasts by machine learning.
\newblock \emph{Journal of Information Processing}, 32:\penalty0 873--880,
  2024.

\bibitem[JMA(2024)]{JMA2024}
JMA.
\newblock Outline of the operational numerical weather prediction at the japan
  meteorological agency.
\newblock
  \url{https://www.jma.go.jp/jma/jma-eng/jma-center/nwp/outline2024-nwp/index.htm},
  2024.
\newblock (Accessed: 22 September 2024).

\bibitem[J{\o}rgensen(1997)]{jorgensen1997}
Bent J{\o}rgensen.
\newblock \emph{The Theory of Dispersion Models}.
\newblock Chapman \& Hall, 1997.

\bibitem[Ke et~al.(2017)Ke, Meng, Finley, Wang, Chen, Ma, Ye, and
  Liu]{ke2017lightgbm}
Guolin Ke, Qi~Meng, Thomas Finley, Taifeng Wang, Wei Chen, Weidong Ma, Qiwei
  Ye, and Tie-Yan Liu.
\newblock Lightgbm: A highly efficient gradient boosting decision tree.
\newblock \emph{Advances in neural information processing systems}, 30, 2017.

\bibitem[Kingma(2014)]{kingma2014adam}
Diederik~P Kingma.
\newblock Adam: A method for stochastic optimization.
\newblock \emph{arXiv preprint arXiv:1412.6980}, 2014.

\bibitem[Kudo(2022)]{kudo2022statistical}
Atsushi Kudo.
\newblock Statistical post-processing for gridded temperature prediction using
  encoder--decoder-based deep convolutional neural networks.
\newblock \emph{Journal of the Meteorological Society of Japan. Ser. II},
  100\penalty0 (1):\penalty0 219--232, 2022.

\bibitem[{LightGBM Contributors}(2026)]{lightgbm_params}
{LightGBM Contributors}.
\newblock Lightgbm parameters.
\newblock \url{https://lightgbm.readthedocs.io/en/latest/Parameters.html},
  2026.
\newblock Accessed: 2026-03-18.

\bibitem[Liu et~al.(2023)Liu, Lou, Yan, Qi, Jin, Yu, Yang, Zhao, and
  Xia]{liu2023deep}
Qi~Liu, Xiao Lou, Zhongwei Yan, Yajie Qi, Yuchao Jin, Shuang Yu, Xiaoliang
  Yang, Deming Zhao, and Jiangjiang Xia.
\newblock Deep-learning post-processing of short-term station precipitation
  based on nwp forecasts.
\newblock \emph{Atmospheric Research}, 295:\penalty0 107032, 2023.

\bibitem[OPTUNA(2024)]{LightGBMTuner}
OPTUNA.
\newblock optuna.integration.lightgbm.lightgbmtuner; optuna 2.0.0
  documentation.
\newblock
  \url{https://optuna.readthedocs.io/en/v2.0.0/reference/generated/optuna.integration.lightgbm.LightGBMTuner.html},
  2024.
\newblock (Accessed: 5 December 2024).

\bibitem[Peng et~al.(2020)Peng, Zhi, Ji, Ji, and Tian]{peng2020prediction}
Ting Peng, Xiefei Zhi, Yan Ji, Luying Ji, and Ye~Tian.
\newblock Prediction skill of extended range 2-m maximum air temperature
  probabilistic forecasts using machine learning post-processing methods.
\newblock \emph{Atmosphere}, 11\penalty0 (8):\penalty0 823, 2020.

\bibitem[RISH(2024)]{RISH2024}
RISH.
\newblock Welcome to rish www data server.
\newblock \url{http://database.rish.kyoto-u.ac.jp/index-e.html}, 2024.
\newblock (Accessed: 22 September 2024).

\bibitem[Rojas-Campos et~al.(2023)Rojas-Campos, Wittenbrink, Nieters,
  Schaffernicht, Keller, and Pipa]{rojas2023postprocessing}
Adrian Rojas-Campos, Martin Wittenbrink, Pascal Nieters, Erik~J Schaffernicht,
  Jan~D Keller, and Gordon Pipa.
\newblock Postprocessing of nwp precipitation forecasts using deep learning.
\newblock \emph{Weather and Forecasting}, 38\penalty0 (3):\penalty0 487--497,
  2023.

\bibitem[Ross(2014)]{ross2014mutual}
Brian~C Ross.
\newblock Mutual information between discrete and continuous data sets.
\newblock \emph{PloS one}, 9\penalty0 (2):\penalty0 e87357, 2014.

\bibitem[Salazar et~al.(2022)Salazar, Che, Zheng, and
  Xiao]{salazar2022multivariable}
Andr{\'e}s~A Salazar, Yuzhang Che, Jiafeng Zheng, and Feng Xiao.
\newblock Multivariable neural network to postprocess short-term, hub-height
  wind forecasts.
\newblock \emph{Energy Science \& Engineering}, 10\penalty0 (7):\penalty0
  2561--2575, 2022.

\bibitem[Shwartz-Ziv and Armon(2022)]{shwartz2022tabular}
Ravid Shwartz-Ziv and Amitai Armon.
\newblock Tabular data: Deep learning is not all you need.
\newblock \emph{Information Fusion}, 81:\penalty0 84--90, 2022.

\bibitem[Tang et~al.(2021)Tang, Ning, Li, Feng, Chen, and
  Xie]{tang2021numerical}
Rongnian Tang, Yuke Ning, Chuang Li, Wen Feng, Youlong Chen, and Xiaofeng Xie.
\newblock Numerical forecast correction of temperature and wind using a
  single-station single-time spatial lightgbm method.
\newblock \emph{Sensors}, 22\penalty0 (1):\penalty0 193, 2021.

\bibitem[TensorFlow(2024)]{TensorFlow}
TensorFlow.
\newblock Tensorflow.
\newblock \url{https://www.tensorflow.org/}, 2024.
\newblock (Accessed: 8 December 2024).

\bibitem[Tsipis et~al.(2023)Tsipis, Banti, Louta, and
  Dimokas]{tsipis2023improving}
Evangelos Tsipis, Konstantina Banti, Malamati Louta, and Nikos Dimokas.
\newblock Improving open weather prediction data accuracy using machine
  learning techniques.
\newblock In \emph{2023 14th International Conference on Information,
  Intelligence, Systems \& Applications (IISA)}, pages 1--8. IEEE, 2023.

\bibitem[Xu et~al.(2020)Xu, Ning, and Luo]{xu2020wind}
Wenqing Xu, Like Ning, and Yong Luo.
\newblock Wind speed forecast based on post-processing of numerical weather
  predictions using a gradient boosting decision tree algorithm.
\newblock \emph{Atmosphere}, 11\penalty0 (7):\penalty0 738, 2020.

\bibitem[Yoshikane and Yoshimura(2022)]{yoshikane2022bias}
Takao Yoshikane and Kei Yoshimura.
\newblock A bias correction method for precipitation through recognizing
  mesoscale precipitation systems corresponding to weather conditions.
\newblock \emph{PLoS Water}, 1\penalty0 (5):\penalty0 e0000016, 2022.

\bibitem[Zhang and Ye(2021)]{zhang2021machine}
Yuhang Zhang and Aizhong Ye.
\newblock Machine learning for precipitation forecasts postprocessing:
  Multimodel comparison and experimental investigation.
\newblock \emph{Journal of Hydrometeorology}, 22\penalty0 (11):\penalty0
  3065--3085, 2021.

\end{thebibliography}

\end{document}